\documentclass[iicol,sn-apa]{sn-jnl}

\usepackage{graphicx}
\usepackage{textcomp}
\usepackage{mathrsfs}
\usepackage{wrapfig}
\usepackage{colortbl}
\usepackage{subfigure}
\usepackage{multirow}

\jyear{2021}%

\theoremstyle{thmstyleone}%
\theoremstyle{thmstyletwo}%

\theoremstyle{thmstylethree}%

\begin{document}

\title[Article Title]{\textit{$\mathcal{A}$$\mathcal{C}$lassi$\mathcal{H}$onk}: A System Framework to Annotate and Classify Vehicular Honk from Road Traffic}


\author*[1]{\fnm{Biswajit} \sur{Maity}}\email{biswajit.maity1@gmail.com}
\author[2]{\fnm{Abdul} \sur{Alim}}\email{aalimv187@gmail.com}
\author[2]{\fnm{Popuri Sree Rama Charan} \sur{}}\email{psr.19U10738@btech.nitdgp.ac.in}
\author[2]{\fnm{Subrata} \sur{Nandi}}\email{subrata.nandi@gmail.com}
\author[2]{\fnm{Sanghita} \sur{Bhattacharjee}}\email{sbhattacharjee.cse@nitdgp.ac.in}


\affil[1]{\orgdiv{Computer Application and Science}, \orgname{Institute of Engineering and Management}, \orgaddress{\postcode{700091}, \state{West Bengal}, \country{India}}}

\affil[2]{\orgdiv{CSE}, \orgname{National Institute and Technology, Durgapur}, \orgaddress{\postcode{713209}, \state{West Bengal}, \country{India}}}


\abstract{
Some recent studies highlight that vehicular traffic and honking contribute to more than 50\% of noise pollution in urban or sub-urban cities in developing regions, including Indian cities. Frequent honking has an adverse effect on health and hampers road safety, the environment, etc. Therefore, recognizing the various vehicle honks and classifying the honk of different vehicles can provide good insights into environmental noise pollution. Moreover, by classifying honks based on vehicle types, we can infer the contextual information of a location, area, or traffic. So far, the researchers have done outdoor sound classification and honk detection, where vehicular honks are collected in a controlled environment or in the absence of ambient noise. Such classification models fail to classify honk based on vehicle types. Therefore, it becomes imperative to design a system that can detect and classify honks of different types of vehicles from which we can infer some contextual information. In this paper, we have developed a novel framework \textit{$\mathcal{A}$$\mathcal{C}$lassi$\mathcal{H}$onk} that performs raw vehicular honk sensing, data labeling and classifies the honk into three major groups, i.e., light-weight vehicles, medium-weight vehicles, and heavy-weight vehicles. We collected the raw audio samples of different vehicular honking based on spatio-temporal characteristics and converted them into spectrogram images. We have proposed a deep learning-based Multi-label Autoencoder model (MAE) for automated labeling of the unlabeled data samples, which provides 97.64\% accuracy in contrast to existing deep learning-based data labeling methods. Further, we have used various pre-trained models, namely Inception V3, ResNet50, MobileNet, ShuffleNet, and proposed an Ensembled Transfer Learning model (EnTL) for vehicle honks classification and performed comparative analysis. Results reveal that EnTL exhibits the best performance compared to pre-trained models and achieves 96.72\% accuracy in our dataset. In addition, we have identified a context of a location based on these classified honk signatures in a city.

}




\maketitle

\section{Introduction}\label{sec1}
The fast growth in urbanization and industrialization has brought about modernization in people's life, but it has several negative effects on urban population. In recent years, noise pollution has emerged as a significant concern, especially in cities of developing economies. Several studies have shown the adverse effects of noise pollution on people’s health and life. For example, prolonged noise exposure can cause hearing problem, sleeping disturbance, cardiovascular diseases \cite{gupta2018noise, jariwala2017noise}. Study \cite{firdaus2010noise} reports that exposure to a high level of noise can lead to the complex health problems, significantly affecting the health. Besides this, factors contributing to urban noise pollution were extensively studied in literature \cite{hammer2014environmental,michali2021noise}. The effects of traffic, construction noise on urban noise rising were explored on the overall urban noise landscape \cite{kalawapudi2020noise} and noise maps has been created to assess noise level for different times throughout a day. Several studies and report have shown that traffic noise is the prime contributor to urban noise. To monitor noise level in urban road traffic, some studies proposed noise maps \cite{chouksey2023heterogeneous,andrade2024urban}. Recently, researchers have examined the pollution level changes in roads before and after COVID-19 lockdown using regression analyses \cite{marwah2022covid}. Despite these advancements, current noise maps and monitoring strategies lack to provide contextual information and sources of noise, which could further assist city planning bodies in devising plans to mitigate noise pollution levels. Notably, vehicular honking is a major source to traffic related noise. Characterizing vehicular honking can provide valuable insights into traffic state characterization, noise estimation, prediction. Hence, it becomes imperative to assess noise pollution, particularly honking, to devise urban traffic planning and enhancing urban sustainability.

However, precise detection methods and classification models of honks corresponding to vehicle types that generates such are missing especially in presence of other outdoor noise sources and heterogeneous mixed model traffic conditions. Existing outdoor noise classification models \cite{sen2010horn,dim2020smartphone,piczak2015environmental,salamon2017deep} fail to identify sub-classification of honk due to lack of annotated data required for model training. In this work, we aim to address the problem of sub-classifying honk source based on vehicle type that emits the honk.
Such sub-classification of honk model can complement the existing noise monitoring \cite{mann2024random}, mapping \cite{andrade2024urban} and modelling research effort \cite{medina2022urban,hu2022comprehensive}. Furthermore, we can develop various micro-services to diminish the exposure to noise pollution due to road traffic for healthy living. Hence, in this research work, we have focused on annotating and classifying vehicular honks of different types of vehicles. Here, we define three types of vehicles based on their sizes: light-weight vehicles (LWV), medium-weight vehicles (MWV), and heavy-weight vehicles (HWV) \cite{shekhar2022liver}. The details of each type of vehicle are described in Table \ref{tab:dv}. Furthermore, we have estimated the context of a location that might help in identifying personal noise exposure and apparently, avoid the highly congested road traffic area based on its spatio-temporal nature. This work has two-fold objectives: \textit{(a) the classification of different vehicular honks in the presence of various ambient noises, (b) characterize the vehicular honks in such a way so that we can infer some meaningful insights from it, which can further be used to mitigate the harmful effects of honking in our daily lives.}

\begin{table}[h]
\centering
\caption{Description of vehicle types}
\begin{tabular}{|c|c|c|c|}
\hline
\rowcolor[HTML]{C0C0C0} 
Features & LWV                                                                                & MWV                                                          & HWV                                                        \\ \hline
Size     & Small                                                                              & Medium                                                       & Large                                                      \\ \hline
Weight   & Light                                                                              & Medium                                                       & Heavy                                                      \\ \hline
Type     & \begin{tabular}[c]{@{}c@{}}Motorbike, Scooty,\\Auto Rickshaw\end{tabular} & \begin{tabular}[c]{@{}c@{}}Cars\end{tabular} & \begin{tabular}[c]{@{}c@{}}Bus, Truck\end{tabular} \\ \hline
\end{tabular}
\label{tab:dv}
\end{table}

\begin{figure}
    \centering
    \includegraphics[width=7cm, height= 5cm]{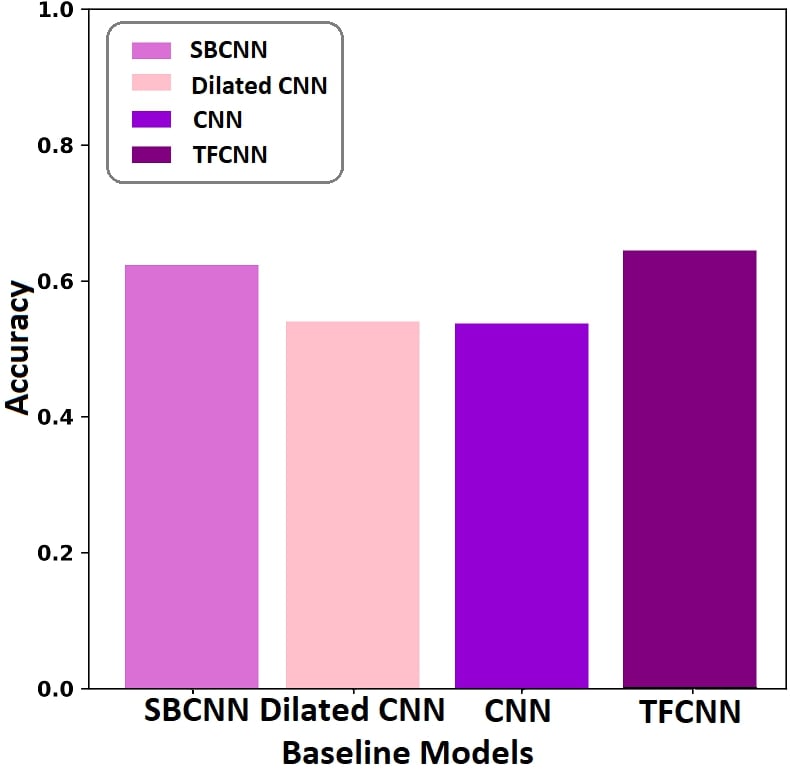}
    \caption{Accuracy of different baseline models with our 30 mins of label data}
    \label{fig:fbl}
\end{figure}

\textbf{Motivation:}
In the last few years, researchers tried to identify vehicular honks by modeling the raw audio signals using Fast Fourier Transformation (FFT), Mel Frequency Cepstral Coefficients (MFCCs), Spectrogram, etc. Additionally, some researchers have developed environmental sound classification techniques (where car honk is considered as one of the classes) to detect and classify environmental sounds \cite{piczak2015environmental,salamon2017deep,zhou2017using}. They have classified the sound of air conditioners, gunfire, street music, automobile horns, children playing, dogs barking, drilling, idling engines, sirens, etc. Different deep-learning models are predominantly used in their works. In this instance, the authors proposed SB-CNN \cite{salamon2017deep}, ConvNet \cite{zhou2017using}, TFCNN \cite{mu2021environmental}, and other CNN model variants. For model training, the aforementioned cited works utilized datasets such as ESC-50, ESC-10, and UrbanSound8K. It is worth noting that ESC-10 does not include any honk sample. However, both ESC-50 and UrbanSound8K contain car honk samples, with duration ranging from 2 seconds to 9 minutes where ESC-50 contains samples with an average duration of 3 minutes, while UrbanSound8K includes samples with an average duration of 134 minutes. Furthermore, it is noted that the majority of these honk samples were taken in a quiet, and controlled environment. Apart from this sound classification, FFT and band-pass filtering-based honk identification are also discussed in \cite{sen2010horn,dim2020smartphone}, and MFCC-based techniques are addressed in \cite{banerjee2012two}. However, they did not deploy any model which can be able to detect honk automatically. In \cite{maity2022dehonk}, the author modeled the honk signals as a spectrogram image and deployed a proper model to detect honk. Furthermore, several applications, like helping hearing-impaired \cite{takeuchi2014smart}, healthier route recommendation system \cite{maity2022dehonk}, are developed based on vehicular honk. As a preliminary experiment, we considered some of the baseline models (SB-CNN \cite{salamon2017deep}, Dilated CNN \cite{chen2019environmental}, CNN \cite{demir2020new}) for honk classification. We used $\sim$ 30 mins of manually labeled samples to perform the same. The accuracy achieved by each model is shown in Fig. \ref{fig:fbl}. It is seen that the accuracy of all the models lies between 57\% to 63\%, which is comparatively much low. For that reason, we need further investigation in terms of model training as well as label data generation. 

\textbf{Issues \& Challenges:}
In this work, we faced several significant challenges. The three main system-level challenges that we overcome to increase honk classification accuracy in presence of background noise are as follows: \textit{First}, the collection of adequate spatio-temporal data: The currently available honk sample datasets (i.e., ESC-50, UrbanSound8k) are insufficient, and were primarily recorded sounds with no background noise. Moreover, these data were not collected with consideration of the spatial and temporal characteristics of a specific location. \textit{Secondly}, Label data creation: When dealing with large volumes of data, manually tagging honk samples from raw traffic audio data can be both laborious and susceptible to errors. Additionally, existing automated data annotation techniques are often not suitable for our case due to the similarity in honk signatures. \textit{Thirdly}, selection of appropriate models for honk classification: Different types of honks exhibit a range of characteristics, including intensity, pitch, duration, volume, etc. It is, therefore, difficult to choose appropriate threshold values that cover all of these groups. The honk characteristics present in recorded signals may be partially masked by ambient noise, such as extraneous speech, loud music, bioacoustic noise, electrical noise, and more. This complexity makes it challenging to accurately identify or filter honks from raw audio samples. Additionally, there may be instances where similarities exist between the patterns of MWV and HWV, further complicating the classification process. Therefore, it is crucial to filtering-based data annotation techniques. Hence, we need to fine-tune deep learning models to develop an automated honk classification system.

\textbf{Contribution:}
In order to effectively classify vehicular honks while there is oncoming traffic, this paper addresses the aforementioned difficulties and introduces a novel framework \textit{$\mathcal{A}$$\mathcal{C}$lassi$\mathcal{H}$onk}. Our contributions encompass the preparation of a substantial volume of datasets, the proposal of a novel model for dataset labeling, and the fine-tuning of a suitable transfer learning model that surpasses the performance of alternative models. The primary contributions of this endeavor are outlined as follows.

\begin{itemize}
    \item Considering significant spatio-temporal variation, we have recorded raw audio samples by two volunteers using our developed Android application AudREC over a 463 km spanning $\sim$13 km road segment in a sub-urban city of Durgapur, India. The details specification is mentioned in Section \ref{sec:da}.
    
    \item Since these raw audio samples can not be directly applied to the deep learning models, proper data modeling and labeling are required beforehand. As manual labeling is cumbersome and error-prone, in this research work, we have designed a \textbf{Multi-label Autoencoder model (MAE)} and \textbf{Multi-label Autoencoder Generative Adversarial Network (MAEGAN)} for labeling the unlabeled samples. Experimental results show that the MAE model performs better than MAEGAN, at around 97.64\% accuracy at the time of data annotation. In total, 54705 samples are labeled, i.e., equivalent to $\sim$15 hours. Furthermore, we have increased the sample set from 54705 to 134286 by using data augmentation techniques. In contrast to the previous UrbanSound8k's 8000 tagged samples, the resulting richly curated dataset of large volume is employed in our experiments, which can serve as an important resource for other researchers (refer Section \ref{sec:dl} and \ref{sec:dac}).
    
    \item As per the literature, the spectrogram is a better choice than FFT/MFCC in honk identification/classification. We have studied the utility of various transfer learning models for the same. In this work, four pre-trained CNN models such as  MobileNet \cite{attallah2023cercan}, ShuffleNet \cite{elhassan2021dsanet}, ResNet 50 \cite{de2022geometric}, and Inception V3 \cite{wang2019pulmonary} are used as transfer learning and their classification performances are investigated thoroughly. Furthermore, We have designed and implemented \textbf{transfer learning-based ensemble method named EnTL} by combining four pre-trained models for the task of vehicular honk classification. Experimental results show that EnTL performs better than other models with 96.72 \% accuracy (refer Section \ref{sec:phc}).
    
    \item The proposed EnTL has been compared with the baseline models like SB-CNN \cite{salamon2017deep}, Dilated CNN \cite{chen2019environmental}, CNN \cite{demir2020new}, and TFCNN \cite{mu2021environmental}. EnTL improves the accuracy $\sim$9\% to 21\%, indicating its effectiveness in honk classification in the presence of ambient noises (refer Section \ref{sec:cbw}).
    
    \item We demonstrate the usefulness of classified honk and sound pressure level (SPL) by detecting outdoor contexts like residential areas, highways, marketplaces, less crowded traffic areas, etc. This inference will help to design several micro-services in the future (refer Section \ref{sec:stc}).

\end{itemize}

The remaining part of the paper is organized as follows: In Section \ref{sec:rw}, we discuss and study existing works. The framework of our proposed system is provided in Section \ref{sec:fw}. The nitty-gritty details of data collection and honk signature analysis are figured out in Section \ref{sec:da}. Section \ref{sec:met} describes our proposed methodology where we briefly describe data labeling, augmentation, and model selection for training. Results are analyzed in Section \ref{sec:ra}. In this section, we also illustrate the procedure to identify the context of a location. Finally, in Section \ref{sec:cf}, we conclude our paper and discuss potential directions for our work.

\section{Related work}
\label{sec:rw}
Noise pollution or disturbance is the outrageous amount of noise that affects the entire ecosystem. The major source of outdoor noise is vehicular honk which is mainly generated in road traffic areas. Due to the nature of our environment, it becomes challenging to reduce noise pollution levels to sustain and maintain the ecosystem. However, we can try to avoid it through a spectacular boom in software technology. In the last few decades, researchers have addressed and adopted several techniques to characterize the noise pollution level in road traffic areas.

\subsection{Noise pollution monitoring \& assessment} Noise level characterization is mainly done by monitoring, forecasting, and identifying the source of noise in the environment. Several techniques (like deploying heterogeneous sensors in different places of cities/smartphone-based monitoring) are adopted to measure the pollution level, where smartphone-based sensing is most popular one. Smartphones are basically used for spatial data collection, while fixed sensor-based devices are deployed for temporal data collection. 

\textbf{\textit{Smartphone-base monitoring:}} In \cite{zipf2020citizen,jezdovic2021crowdsensing}, the author collected the data using smartphones and notified the efficacious urban planning corresponding to noise pollution exposure. Spatio-temporal patterns of noise pollution are also measured using smartphones in \cite{zamora2017accurate}. Apart from this, the researchers in \cite{maity2022coan, allen2009spatial} have addressed a correlation between noise and air pollution to recognize the primary sources of environmental noise clearly. Due to the sensitivity issue of the smartphone and the lack of proper calibration techniques, infrastructure-based sensing seems more realistic.

\textbf{\textit{Fixed sensor-based monitoring:}}  To get the temporal variation of data, different calibrated sensors can be deployed throughout the city (mainly in the road traffic area) to measure the noise pollution level. In previous studies, the researchers developed a wireless sensor network \cite{santini2008first} in order to monitor noise pollution. However, nowadays, many advanced sensors are developed that can capture data with a high level of accuracy \cite{bello2019sonyc}. Raspberry Pi-controlled cloud-based sensor is also designed in \cite{saha2018raspberry} to do the same. Moreover, an Arduino controller with IoT technology \cite{ezhilarasi2017system} is also used as a fixed sensing strategy to get the noise data from the environment. Like \cite{santini2008first}, an advanced wireless sensor-based system is designed for the same in \cite{segura2014low}. Besides the advancement of technology, some pros exist that make the sensors unreliable. The sensors may provide erroneous data or stop functioning due to environmental hazards such as storms, rainfall, etc.

\textbf{\textit{Noise level forecasting:}} To predict the noise pollution level in road traffic, the researchers \cite{medina2022urban} have developed several models and techniques. In \cite{garg2015applications, guarnaccia2017development}, the authors proposed ARIMA model-based time series traffic noise prediction model. Deep learning-based models are also used in several kinds of research to forecast pollution levels. In \cite{navarro2020sound}, the authors developed LSTM-based models to predict sound pollution levels in urban road traffic. In our previous work \cite{maity2022predhonk}, we also estimated the noise pollution level in the road traffic area by predicting the vehicular honk count using the E-LSTM model.

\textbf{\textit{Noise source identification:}} Apart from the above mentioned techniques, noise pollution measurement by identifying the source is another way to monitor pollution. In \cite{vera2018towards, suvorov2018deep}, the author identified the source of sound using a deep neural network. Sound source identification using a microphone array configuration is another effective technique to measure pollution exposure. In \cite{grondin2019lightweight}, the author used an open and closed microphone array for sound localization. Wireless Acoustic Sensor Networks-based sound source identification is addressed in \cite{cobos2017survey} by the authors.

\subsection{Vehicular honk identification \& applications} The primary source of noise pollution is traffic noise, which is mainly comprised of different vehicular honks. Therefore, vehicular honk identification and classification of the different types of vehicle honk are essential to characterize the noise pollution level. Several researchers are already identified vehicular honks by adopting different strategies. These strategies are mainly classified into three categories: a) classifying environmental sound, where vehicular honk is one of the classes, b) determining the honks directly from the audio samples by using FFT, Spectrogram, or MFCC, and c) applications based on the honk signature. Details of the honk identification and sound classification methods are summarized in Table \ref{tab:lr}.

\textbf{\textit{Environmental sound classification:}}
There are numerous works that identify vehicular honking by classifying environmental sounds, where car honking is one of the class \cite{piczak2015environmental,salamon2017deep,zhou2017using,khamparia2019sound,abdoli2019end,mesaros2019sound,chen2019environmental,demir2020new,ahmed2020automatic,mushtaq2020environmental,guzhov2021esresnet,mu2021environmental}. Authors mainly used ESC-10, ESC-50, or UrbanSound8K dataset for sound classification. All the data are collected in a controlled environment where ambient sound is not present. Most of the cases, they modified the CNN models to increase the accuracy. The highest level of accuracy that they have achieved till yet is 97\% for the ESC-10 dataset in \cite{guzhov2021esresnet}. However, the ESC-10 data set does not contain the honk class. Apart from ESC-10, the maximum accuracy achieved by the DCASE-2017 ASC dataset \cite{demir2020new} is 96.23\%, which contains a car honk as a class. 

\textbf{\textit{Honk from raw samples and its applications:}}
In \cite{sen2010horn}, the author used band-pass filtering to remove the noises from the raw audio samples and then applied FFT to detect the honks. They also calculated the duration of each honk in this work. Nevertheless, their selected threshold values for band-pass filters don't guarantee a definite honk signature all the time. Furthermore, any framework/system was not designed that can identify honk automatically while moving. A similar kind of FFT-based honk detection technique is found in \cite{dim2020smartphone}. As an application of honk, the authors developed a system that assists the hearing-impaired person in driving. In another work \cite{takeuchi2014smart}, a smartphone-based system is implemented to detect honking and then generate alarm sounds for hearing-impaired people. Apart from these, an embedded system is designed to identify emergency honk \cite{palecek2016emergency}, and MFCC-based honk detection in \cite{banerjee2012two}. In our previous work \cite{maity2022dehonk}, we have identified vehicular honk from raw audio samples in the presence of several ambient noises. Apart from this, we have also determined different features of the honk, like duration of the honk, the inter-honk gap between two successive honks, and honk count, etc., and based on these features, we have tried to identify the context of a location and recommend a honk-aware route, which is a healthier route in comparison to the google recommended route. 

As per the works mentioned above, it is apparent that the researchers are working on honk identification and environmental sound classification techniques. \textit{However, fine-grained vehicular honk classification in terms of vehicle types is our primary objective of the work presented in this paper, and it is not yet done.}

\begin{sidewaystable*}[htpb]
\footnotesize
\centering
\caption{Survey for vehicular honk identification and environmental sound classification, where vehicular honk is considered as a class }
\begin{tabular}{|ccccc|}
\hline
\rowcolor[HTML]{C0C0C0} 
\multicolumn{1}{|c|}{\cellcolor[HTML]{C0C0C0}\textbf{Author, Year}}                       & \multicolumn{1}{c|}{\cellcolor[HTML]{C0C0C0}\textbf{Dataset}}                                            & \multicolumn{1}{c|}{\cellcolor[HTML]{C0C0C0}\textbf{Technique}}                                                                                                                             & \multicolumn{1}{c|}{\cellcolor[HTML]{C0C0C0}\textbf{Accuracy Obtained}}                                                                           & \textbf{Remarks}                                                                                                                                                  \\ \hline
\multicolumn{5}{|c|}{\textbf{Honk Detection from raw audio samples}}                                                                                                                                                                                                                                                                                                                                                                                                                                                                                                                                                                                                                                       \\ \hline
\multicolumn{1}{|c|}{\begin{tabular}[c]{@{}c@{}}R. Sen et al.\\ (2010) \cite{sen2010horn}\end{tabular}}      & \multicolumn{1}{c|}{Self collected}                                                                      & \multicolumn{1}{c|}{FFT}                                                                                                                                                                    & \multicolumn{1}{c|}{\begin{tabular}[c]{@{}c@{}}Only identify the honks, \\ They have not trained any model\end{tabular}}                          & \begin{tabular}[c]{@{}c@{}}No ML/DL \\ is used\end{tabular}                                                                                                       \\ \hline
\multicolumn{1}{|c|}{\begin{tabular}[c]{@{}c@{}}R. Banerjee et al.\\ (2012) \cite{banerjee2012two}\end{tabular}} & \multicolumn{1}{c|}{Self collected}                                                                      & \multicolumn{1}{c|}{Modified MFCC}                                                                                                                                                          & \multicolumn{1}{c|}{\begin{tabular}[c]{@{}c@{}}Only identify the honks, \\ They have not trained any model\end{tabular}}                          & \begin{tabular}[c]{@{}c@{}}No ML/DL \\ is used\end{tabular}                                                                                                       \\ \hline
\multicolumn{1}{|c|}{\begin{tabular}[c]{@{}c@{}}Takeuchi et al.\\ (2014) \cite{takeuchi2014smart}\end{tabular}}    & \multicolumn{1}{c|}{---}                                                                                 & \multicolumn{1}{c|}{IIR Comb Filter}                                                                                                                                                        & \multicolumn{1}{c|}{---}                                                                                                                          & \begin{tabular}[c]{@{}c@{}}Develop an application for \\ hearing impaired people\end{tabular}                                                                     \\ \hline
\multicolumn{1}{|c|}{\begin{tabular}[c]{@{}c@{}}Josef et al.\\ (2016) \cite{palecek2016emergency}\end{tabular}}       & \multicolumn{1}{c|}{---}                                                                                 & \multicolumn{1}{c|}{Embedded system}                                                                                                                                                        & \multicolumn{1}{c|}{---}                                                                                                                          & \begin{tabular}[c]{@{}c@{}}Own develop hardware \\ system to detect honk\end{tabular}                                                                             \\ \hline
\multicolumn{1}{|c|}{\begin{tabular}[c]{@{}c@{}}Dim et al.\\ (2020 \cite{dim2020smartphone}\end{tabular}}         & \multicolumn{1}{c|}{---}                                                                                 & \multicolumn{1}{c|}{\begin{tabular}[c]{@{}c@{}}FFT and self design honk \\ detection algorithm\end{tabular}}                                                                                & \multicolumn{1}{c|}{---}                                                                                                                          & \begin{tabular}[c]{@{}c@{}}Develop an application for \\ hearing impaired people\end{tabular}                                                                     \\ \hline
\multicolumn{1}{|c|}{\begin{tabular}[c]{@{}c@{}}Biswajit et al.\\ (2022) \cite{maity2022dehonk}\end{tabular}}    & \multicolumn{1}{c|}{Self Collected}                                                                      & \multicolumn{1}{c|}{\begin{tabular}[c]{@{}c@{}}Spectrogram to train Different Transfer \\ Learning models\end{tabular}}                                                                     & \multicolumn{1}{c|}{97.69\%}                                                                                                                      & \begin{tabular}[c]{@{}c@{}}Detected honk in presence \\ of ambient noise\end{tabular}                                                                             \\ \hline
\multicolumn{5}{|c|}{\textbf{Sound Classification}}                                                                                                                                                                                                                                                                                                                                                                                                                                                                                                                                                                                                                                                                        \\ \hline
\multicolumn{1}{|c|}{\begin{tabular}[c]{@{}c@{}}Piczak et al.\\ (2015) \cite{piczak2015environmental}\end{tabular}}      & \multicolumn{1}{c|}{\begin{tabular}[c]{@{}c@{}}ESC-50, \\ ESC-10, and\\ UrbanSound8K\end{tabular}}       & \multicolumn{1}{c|}{\begin{tabular}[c]{@{}c@{}}Log-scaled\\ mel-spectrograms and develop\\  a CNN model\end{tabular}}                                                                       & \multicolumn{1}{c|}{\begin{tabular}[c]{@{}c@{}}Best accuracy 87\% \\ for Esc-10 dataset and remaining\\ are less than that\end{tabular}}          & \begin{tabular}[c]{@{}c@{}}10 different class, \\ where car honk is one class\end{tabular}                                                                        \\ \hline
\multicolumn{1}{|c|}{\begin{tabular}[c]{@{}c@{}}Salamon et al.\\ (2017) \cite{salamon2017deep}\end{tabular}}     & \multicolumn{1}{c|}{UrbanSound8K}                                                                        & \multicolumn{1}{c|}{\begin{tabular}[c]{@{}c@{}}Log-scaled\\ mel-spectrograms and \\ develop a SB-CNN model\end{tabular}}                                                                    & \multicolumn{1}{c|}{85\%}                                                                                                                         & \begin{tabular}[c]{@{}c@{}}10 different class, \\ where car honk is one class\end{tabular}                                                                        \\ \hline
\multicolumn{1}{|c|}{\begin{tabular}[c]{@{}c@{}}Zhou et al.\\ (2017) \cite{zhou2017using}\end{tabular}}        & \multicolumn{1}{c|}{UrbanSound8K}                                                                        & \multicolumn{1}{c|}{\begin{tabular}[c]{@{}c@{}}2D mel-spectrogram and \\ develop a ConvNet model\end{tabular}}                                                                              & \multicolumn{1}{c|}{Best accuracy 86\%}                                                                                                           & \begin{tabular}[c]{@{}c@{}}10 different class, \\ where car honk is one class\end{tabular}                                                                        \\ \hline
\multicolumn{1}{|c|}{\begin{tabular}[c]{@{}c@{}}Khamparia et al.\\ (2019) \cite{khamparia2019sound}\end{tabular}}   & \multicolumn{1}{c|}{\begin{tabular}[c]{@{}c@{}}ESC-10 and \\ ESC-50\end{tabular}}                        & \multicolumn{1}{c|}{\begin{tabular}[c]{@{}c@{}}Spectrogram and develop a CNN,\\ tensor deep stacking network (TDSN)\end{tabular}}                                                           & \multicolumn{1}{c|}{\begin{tabular}[c]{@{}c@{}}Proposed CNN 77\% and \\ proposed TDSN 56\%\end{tabular}}                                          & \begin{tabular}[c]{@{}c@{}}10 different class, \\ where car honk is included\end{tabular}                                                                        \\ \hline
\multicolumn{1}{|c|}{\begin{tabular}[c]{@{}c@{}}Sajjad et al.\\ (2019) \cite{abdoli2019end}\end{tabular}}      & \multicolumn{1}{c|}{UrbanSound8K}                                                                        & \multicolumn{1}{c|}{\begin{tabular}[c]{@{}c@{}}Audio signals of any length as it \\ splits the signal into overlapped \\ frames using a sliding window, \\ and used1D CNN\end{tabular}}     & \multicolumn{1}{c|}{89\%}                                                                                                                         & \begin{tabular}[c]{@{}c@{}}10 different class, \\ where car honk not included\end{tabular}                                                                        \\ \hline
\multicolumn{1}{|c|}{\begin{tabular}[c]{@{}c@{}}Mesaros et al.\\ (2019) \cite{mesaros2019sound}\end{tabular}}     & \multicolumn{1}{c|}{\begin{tabular}[c]{@{}c@{}}freesound.org,\\ TUT Sound, and \\ AudioSet\end{tabular}} & \multicolumn{1}{c|}{\begin{tabular}[c]{@{}c@{}}log-mel energies, \\ MFCC, pitch, \\ spectrogram and used CNN, DNN,\\ RNN, CRNN, LSTM, GRU\end{tabular}}                                     & \multicolumn{1}{c|}{\begin{tabular}[c]{@{}c@{}}Among all the techniques \\ 95\% is the highest accuracy\end{tabular}}                             & \begin{tabular}[c]{@{}c@{}}Classified 4 categories \\ of sound in \\ three different phase, where \\ car honk is one class\end{tabular}                           \\ \hline
\multicolumn{1}{|c|}{\begin{tabular}[c]{@{}c@{}}Yan et al.\\ (2019) \cite{chen2019environmental}\end{tabular}}         & \multicolumn{1}{c|}{UrbanSound8K}                                                                        & \multicolumn{1}{c|}{\begin{tabular}[c]{@{}c@{}}Log scale mel-spectrogram \\ and proposed Dilated Convolution\end{tabular}}                                                                  & \multicolumn{1}{c|}{78\%}                                                                                                                         & \begin{tabular}[c]{@{}c@{}}10 different class, \\ where car honk is one class\end{tabular}                                                                        \\ \hline
\multicolumn{1}{|c|}{\begin{tabular}[c]{@{}c@{}}Demir et al.\\ (2020) \cite{demir2020new}\end{tabular}}       & \multicolumn{1}{c|}{\begin{tabular}[c]{@{}c@{}}DCASE-2017 ASC \\ and UrbanSound8K\end{tabular}}          & \multicolumn{1}{c|}{\begin{tabular}[c]{@{}c@{}}Spectrogram, and used\\  Transfer learning models\end{tabular}}                                                                              & \multicolumn{1}{c|}{\begin{tabular}[c]{@{}c@{}}For DCASE-2017 ASC dataset \\ accuracy is 96.23\% and \\ for UrbanSound8K is 86.70\%\end{tabular}} & \begin{tabular}[c]{@{}c@{}}For 1st dataset 15 category \\ and 2nd dataset 10 category \\ classification is done, \\ by considering honk is one class\end{tabular} \\ \hline
\multicolumn{1}{|c|}{\begin{tabular}[c]{@{}c@{}}Ahmed et al.\\ (2020) \cite{ahmed2020automatic}\end{tabular}}       & \multicolumn{1}{c|}{\begin{tabular}[c]{@{}c@{}}ESC-50, ESC-10,\\ and UrbanSound8K\end{tabular}}          & \multicolumn{1}{c|}{\begin{tabular}[c]{@{}c@{}}Log-scaled\\ mel-spectrograms, \\ and develop a CNN model\end{tabular}}                                                                      & \multicolumn{1}{c|}{\begin{tabular}[c]{@{}c@{}}Highest accuracy achieved 92.9\%\\ among all the dataset\end{tabular}}                             & \begin{tabular}[c]{@{}c@{}}10 different class, \\ where car honk is one class\end{tabular}                                                                        \\ \hline
\multicolumn{1}{|c|}{\begin{tabular}[c]{@{}c@{}}Zohaib et al.\\ (2020) \cite{mushtaq2020environmental}\end{tabular}}      & \multicolumn{1}{c|}{\begin{tabular}[c]{@{}c@{}}ESC-10, ESC-50,\\ and UrbanSound8K\end{tabular}}          & \multicolumn{1}{c|}{\begin{tabular}[c]{@{}c@{}}Mel spectrogram (Mel), Mel Frequency \\ Cepstral Coefficient (MFCC) and Log-Mel\\ and proposed DCNN\end{tabular}}                            & \multicolumn{1}{l|}{\begin{tabular}[c]{@{}l@{}}94.94\%, 89.28\%, and 95.37\% for\\ the respective dataset mentioned here\end{tabular}}            & \begin{tabular}[c]{@{}c@{}}10 different class, \\ where car honk is one class\end{tabular}                                                                        \\ \hline
\multicolumn{1}{|c|}{\begin{tabular}[c]{@{}c@{}}Andrey et al. \\ (2021) \cite{guzhov2021esresnet}\end{tabular}}     & \multicolumn{1}{c|}{\begin{tabular}[c]{@{}c@{}}ESC-50, ESC-10,\\ and UrbanSound8K\end{tabular}}          & \multicolumn{1}{c|}{\begin{tabular}[c]{@{}c@{}}log-power Short Time Fourier \\ Transform (STFT) and used \\ ESRes-Net Attention with and \\ without ImageNet pretrained model\end{tabular}} & \multicolumn{1}{c|}{\begin{tabular}[c]{@{}c@{}}97.0\% (ESC-10), 91.5\% (ESC-50\\ and 84.2\% / 85.4\% \\ (US8K mono / stereo).\end{tabular}}       & \begin{tabular}[c]{@{}c@{}}10 different class, \\ where car honk is one class\end{tabular}                                                                        \\ \hline
\multicolumn{1}{|c|}{\begin{tabular}[c]{@{}c@{}}Wenjie et al.\\ (2021) \cite{mu2021environmental}\end{tabular}}      & \multicolumn{1}{c|}{\begin{tabular}[c]{@{}c@{}}ESC-50, and \\ UrbanSound8K\end{tabular}}                 & \multicolumn{1}{c|}{\begin{tabular}[c]{@{}c@{}}Harmonic spectrogram,\\ Percussive spectrogram, \\ Develop a TFCNN model\end{tabular}}                                                       & \multicolumn{1}{c|}{\begin{tabular}[c]{@{}c@{}}For ESC-50 dataset \\ accuracy is 84.4\% and \\ for UrbanSound8K is 91.3\%\end{tabular}}           & \begin{tabular}[c]{@{}c@{}}10 different class, \\ where car honk is one class\end{tabular}                                                                        \\ \hline
\end{tabular}
\label{tab:lr}
\end{sidewaystable*}

\section{Framework} \label{sec:fw}

\begin{figure*}
    \centering
    \includegraphics[width= 15cm, height=4cm]{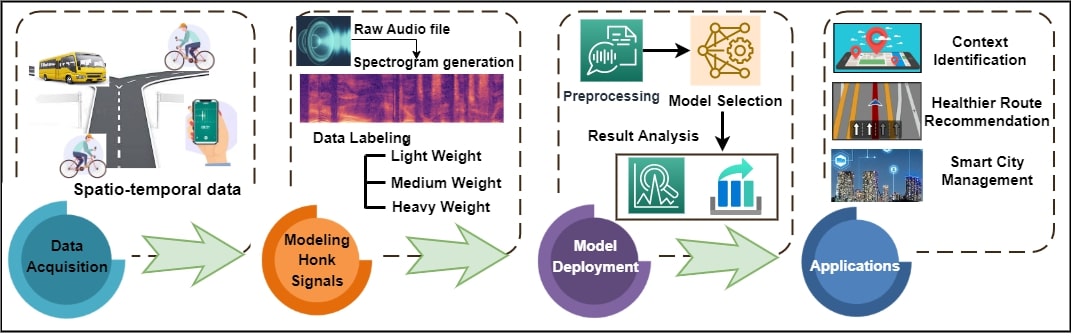}
    \caption{Framework for our proposed system (\textit{$\mathcal{A}$$\mathcal{C}$lassi$\mathcal{H}$onk})}
    \label{fig:fw}
\end{figure*}

Our proposed framework \textit{$\mathcal{A}$$\mathcal{C}$lassi$\mathcal{H}$onk} has four distinct modules as displayed in Fig. \ref{fig:fw}:
\begin{enumerate}[(1)]
\item Data Acquisition: Data collection is one the most crucial to our work since it allows us to accomplish our tasks. For collecting honk data of different vehicles, we have developed a customized android based application. Further, we process these data for honk classification.

\item Modeling Honk Signals: Raw audio samples are converted to spectrogram images. These spectrogram images are labeled by our proposed Multi-label autoencoder model (MAE). In addition, augmentation techniques are used to generate the synthetic data, which is used to increase the data volume.

\item Model Deployment: Different convolutional neural network (CNN) are used to classify the honk signature. The results of each model are analyzed, and the best-suited models are selected for classification.

\item Applications: Depending upon the nature of honking and the number of particular types of vehicles in a specific area, a unique pattern can be identified, which is further used to determine the context of a location. Once the context is identified, then various applications can be developed.

\end{enumerate}

\section{Data acquisition} \label{sec:da}
Our motivation is to identify the context of a location based on different vehicle honking, and sound pressure levels (SPL) by considering spatio-temporal characteristics. To do that, we need a wide range of data collection throughout the city in different places like residential areas, less/high traffic areas, marketplaces, schools/colleges, etc. In this research, a customized android application is developed to collect different vehicular honking data. The application is GPS enabled and it can store the data in the form of raw audio samples as well as it can generate a text file that contains the timestamp, SPL level, and intensity value along with the location. At the time of the data collection, we used 8kHz as a sample rate, a bit depth of 16bits, mono as an audio channel, and an audio format as Wave(.wav). We have mainly obscured two different routes of Durgapur, a sub-urban city in India. Intuitively, we choose each route so that we can cover different demographic areas, which will help us to determine the context of a location. To observe the temporal variation of vehicular honking, we divided each day into three segments: morning, afternoon, and evening, and collected data for each route during these time periods. We traversed 3.5 km to 23 km in each time slot, totaling approximately 463 km distance covered. As transportation, modes, buses were utilized to cover longer distances, while cycles were used to cover shorter distances. We collected around 15 hours of data as a whole. Details of data collection information are represented in Table \ref{tab:dc}. 

\begin{table*}[]
\footnotesize
\centering
\caption{Details representation of data collection procedure }
\begin{tabular}{|l|c|c|c|c|l|}
\hline
\rowcolor[HTML]{C0C0C0} 
\multicolumn{1}{|c|}{\cellcolor[HTML]{C0C0C0}Route \#} & \begin{tabular}[c]{@{}c@{}}Distance covered\\ in each trail\end{tabular} & \begin{tabular}[c]{@{}c@{}}Total distance\\ covered\end{tabular} & \begin{tabular}[c]{@{}c@{}}Audio data \\ duration\end{tabular} & \begin{tabular}[c]{@{}c@{}}Mode of \\ Transport\end{tabular} & \multicolumn{1}{c|}{\cellcolor[HTML]{C0C0C0}Demographic areas}                                                                     \\ \hline
Route 1                                                & 23km                                                                     & 423km                                                            & 712 mins   &Bus                                           & \begin{tabular}[c]{@{}l@{}}Hospital area, Marketplace, Bus terminals,\\ less/high traffic area, educational institute\end{tabular} \\ \hline
Route 2                                                & 3.5km                                                                    & 40.5km                                                           & 202 mins                                                       & Cycle                                                        & \begin{tabular}[c]{@{}l@{}}Less/high traffic area, Marketplace, \\ Residential area\end{tabular}                                   \\ \hline
\end{tabular}
\label{tab:dc}
\end{table*}

\subsection{Spatio-temporal honk signature analysis}
In this section, different vehicle honking features are analyzed by considering spatio-temporal characteristics. We observed 5 minutes of data for three different places (residential area, marketplace, high traffic) and verified the count of honks, SPL level, and correlation between them for each type of vehicle. Moreover, we performed manual data labeling for this experiment. 

\subsubsection{Honk count of different vehicles}
\begin{figure*}
\centering
\subfigure[]{\label{fig:hc1}{\includegraphics[width=5cm, height= 4cm]{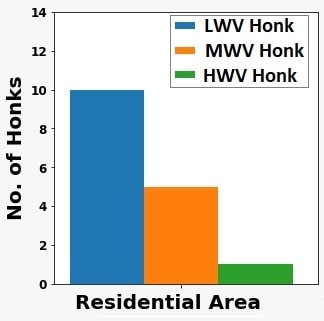}}}
\subfigure[]{\label{fig:hc2}{\includegraphics[width=5cm, height= 4cm]{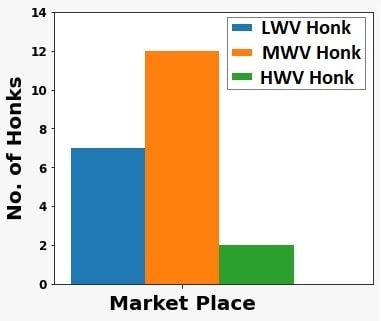}}}
\subfigure[]{\label{fig:hc3}{\includegraphics[width=5cm, height= 4cm]{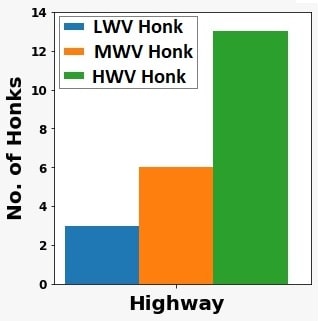}}}

\caption{The number of honks initiate by different types of vehicles in (a) Residential area, (b) Marketplace, and (c) Highway}
\label{fig:hc}
\end{figure*}

The honk frequency produced by different types of vehicles in a specific location can be a crucial parameter to determine the context of a location as well as the pollution level in the environment. To illustrate this, a sample case study was conducted where three different places are considered, such as a residential area, high way, and marketplace, and results are shown in Fig. \ref{fig:hc}. Fig. \ref{fig:hc1} shows that a higher number of honk counts are generated by LWV in the residential area, whereas in the case of the marketplace and highway, MWV and HWV generates a higher number of honk count, respectively, which is depicted in Fig. \ref{fig:hc2} \& \ref{fig:hc3} respectively. A very sensible rhythm is observed in the honk count, which helps us to reach our aim.

\subsubsection{Distribution of sound pressure level}
\begin{figure*}
\centering
\subfigure[]{\label{fig:spl1}{\includegraphics[width=5cm, height= 3.7cm]{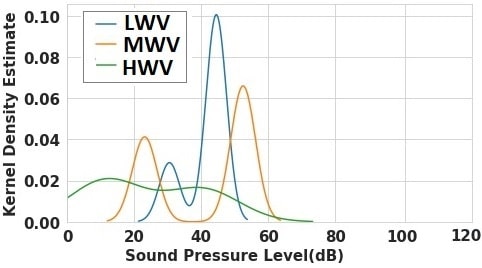}}}
\subfigure[]{\label{fig:spl2}{\includegraphics[width=5cm, height= 3.7cm]{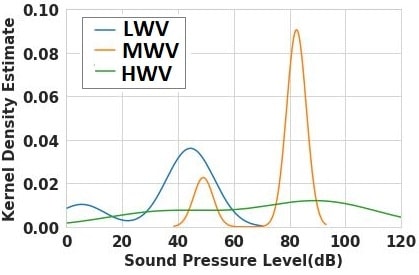}}}
\subfigure[]{\label{fig:spl3}{\includegraphics[width=5cm, height= 3.7cm]{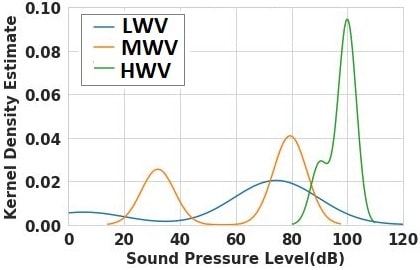}}}

\caption{The distribution of SPL in (a) Residential area, (b) Marketplace, and (c) Highway}
\label{fig:spl}
\end{figure*}

Sound pressure level (SPL), measured in decibel(dB) value, is another important parameter in this study. Different vehicles produce different level dB values. Hence, SPL determination based on the vehicular type is essential and Kerner density function(KDE) is used to represent the distribution of the SPL for a specific area. Fig. \ref{fig:spl} represents the distribution of the  SPL for three different locations. We can visualize that the SPL distribution of the LWV is high, and HWV tends to be negligible in a residential area(Fig. \ref{fig:spl1}). In Fig. \ref{fig:spl2}, it is also clearly observed the highest SPL distribution of MWV in the marketplace, whereas in Fig. \ref{fig:spl3}, HWV depicts the highest distribution in Highway. Moreover, it is apparent that the SPL value is in the higher range in Fig. \ref{fig:spl2} and \ref{fig:spl3} in comparison to Fig. \ref{fig:spl1}. Therefore, we can say that there is a spatial variation in the SPL, which might help to determine a context of a location. 

\subsubsection{Correlation between SPL with honk count }
\begin{figure*}
\centering
\subfigure[]{\label{fig:co1}{\includegraphics[width=5cm, height= 4cm]{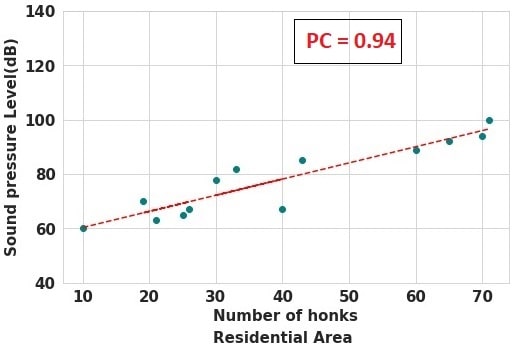}}}
\subfigure[]{\label{fig:co2}{\includegraphics[width=5cm, height= 4cm]{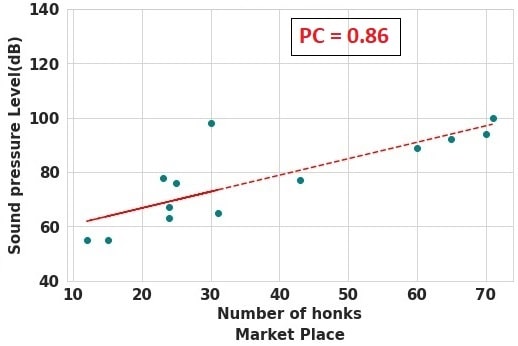}}}
\subfigure[]{\label{fig:co3}{\includegraphics[width=5cm, height= 4cm]{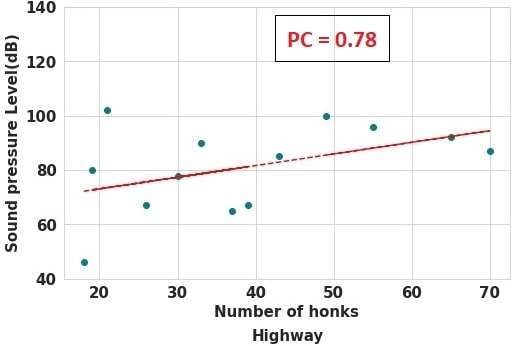}}}

\caption{Correlation between honk count and SPL, in (a) Residential area, (b) Marketplace, and (c) Highway}
\label{fig:co}
\end{figure*}

To estimate the correlation between honk count with the SPL, we have used the well-known Pearson correlation (PC). In Fig. \ref{fig:co}, we notice that in all the locations, PC values signify a strong positive correlation, implying a higher honk count and a higher SPL. Among all the locations, residential areas contain the highest positive correlation (PC= 0.94, see Fig. \ref{fig:co1}) because the traffic flow pattern is almost similar throughout the day in residential areas. In case of a highway, the correlation is strongly positive, but the value of the PC (PC= 0.78,  see Fig. \ref{fig:co2}) is less than in residential areas because the traffic/vehicle movement is not well distributed throughout the day. The nature of vehicle movement is more or less similar to a marketplace, and that is the reason for the PC value of 0.86 (Fig. \ref{fig:co3}), indicating a strong positive correlation.

The experiment outlined above demonstrates a significant correlation between the spatio-temporal characteristics of honks emitted by different vehicles across various scenarios. Manual analysis of these characteristics is inherently challenging and laborious. Consequently, it is imperative to devise a system capable of automatically extracting these properties and accurately classifying honks from unprocessed raw noise samples.

\section{Proposed Methodology} \label{sec:met}
This section details the honk classification process and the selection of its features. Initially, we introduce honk signal modeling, followed by a description of data augmentation, data labeling, and preprocessing techniques. Subsequently, we present various deep learning models for classifying vehicular honks and determine the most suitable model for our proposed method.

\subsection{Spectrogram generation}
\begin{figure*}
\centering
\subfigure[]{\label{fig:spe1}{\includegraphics[width=10cm, height= 6cm]{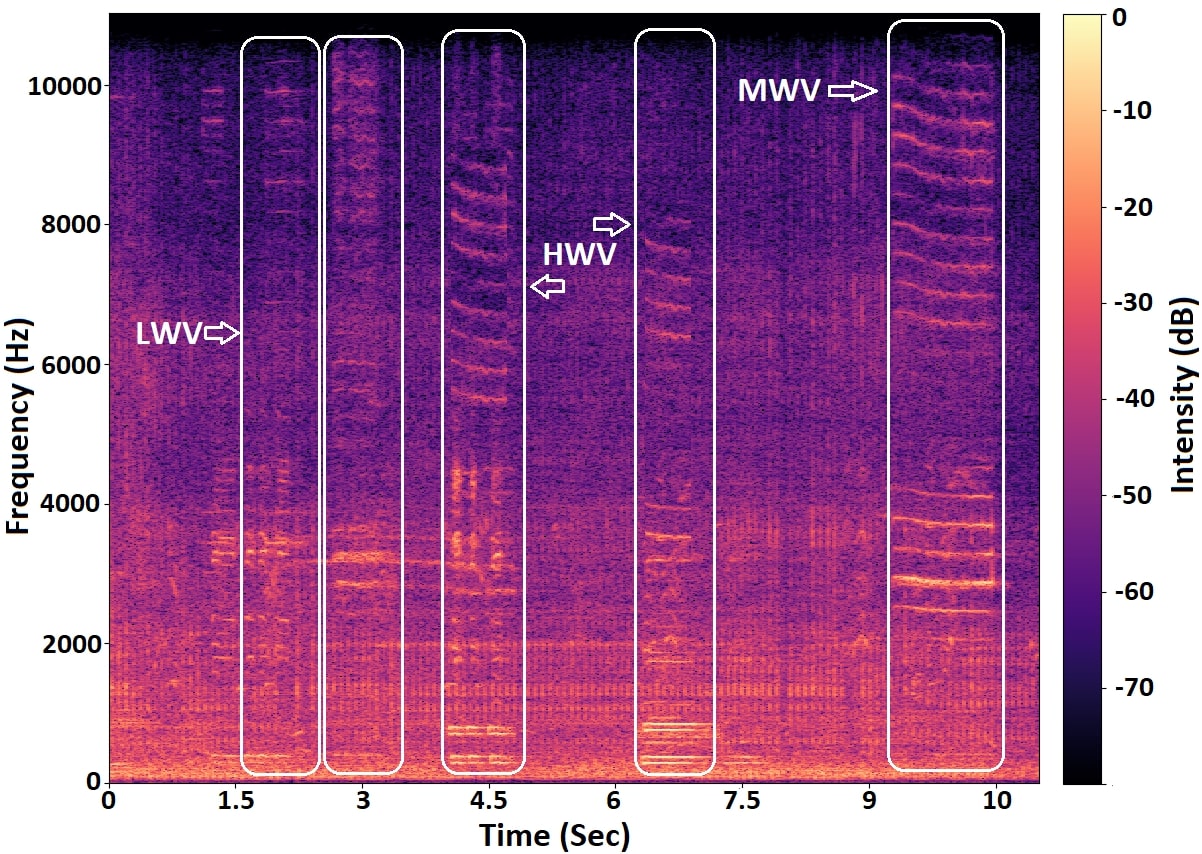}}}\\
\subfigure[]{\label{fig:spe2}{\includegraphics[width=5cm, height= 4cm]{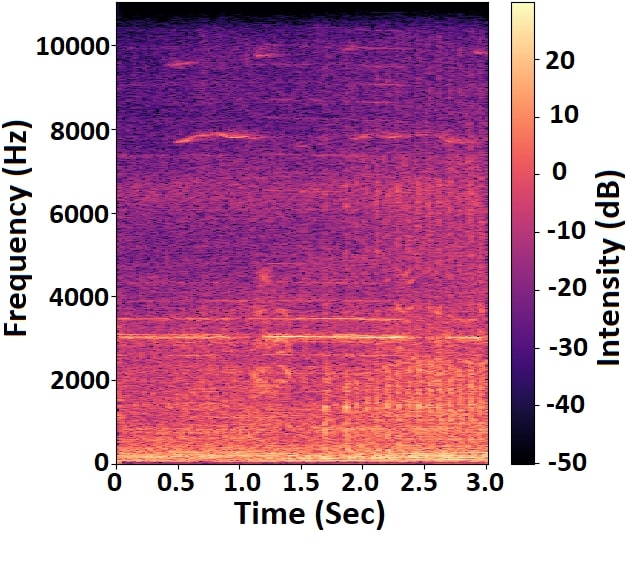}}}
\subfigure[]{\label{fig:spe3}{\includegraphics[width=5cm, height= 4cm]{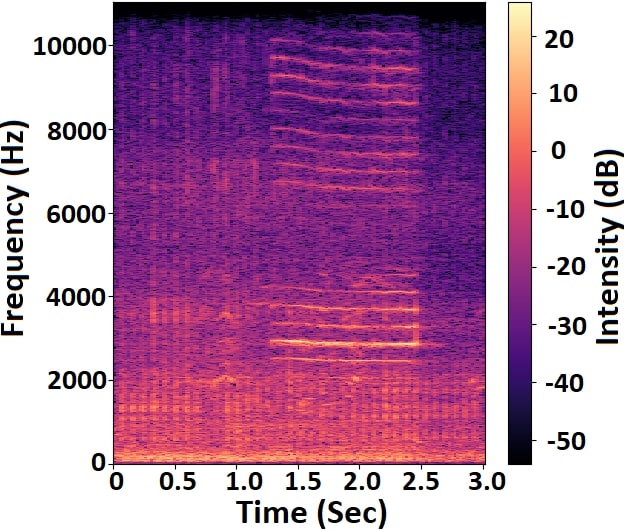}}}
\subfigure[]{\label{fig:spe4}{\includegraphics[width=5cm, height= 4cm]{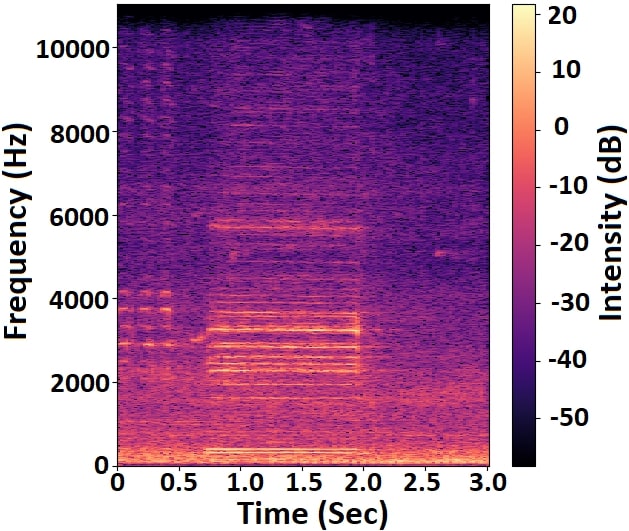}}}

\caption{Sample spectrogram of a road traffic area in absence of ambient noise that represents (a) combinations of different honks (i.e., LWV, MWV, and HWV) in 10-sec duration (b) only LWV (c) only MWV, (d) only HWV  }
\label{fig:spe}
\end{figure*}

A spectrogram is a visual representation of a raw audio file, depicting the loudness of a signal over time at different frequencies present in a waveform. Similar to our previous work \cite{maity2022dehonk}, we used spectrogram to determine honks from raw audio samples. To generate a spectrogram image, we perform Fourier Transformation, which decomposes the signals into frequencies and shows the amplitude of each frequency over time. Like \cite{maity2022dehonk}, the current work generates spectrograms by dividing each audio sample into a 1-seconds segment, with the X-axis representing the time(sec), the Y-axis denoting frequency(Hz), and the amplitude of each frequency is represented by the different colors. Brighter color indicates a higher amplitude. As an example, we have considered a random 10-sec audio clip and converted it into spectrogram images, which is shown in Fig. \ref{fig:spe1}. The figure shows that there are five possible honks (2 LWV, 1 MWV, and 2 HWV) within this 10-sec duration. Furthermore, Fig. \ref{fig:spe2}, \ref{fig:spe3}, and \ref{fig:spe4} illustrates the individual spectrogram image of audio data collected from LWV, MWV, and HWV environment for better understanding. 

\subsection{Data labeling} \label{sec:dl}
Deep learning classifiers require a significant volume of good-quality labeled data. In the absence of such a rich dataset related to the problem presented in this paper, we initially considered manual labeling. Nevertheless, manual labeling is simple, but it is labor-intensive, time-consuming, and more prone to human error. To deal with that, we employed two personnel to label the collected data. In our work, the non-honks data are labeled as '0', LWV as '1'', MWV as '2', and HWV as '3'. Besides audio files, corresponding video files are also used to make our annotation error-free for manual labeling. A total of 1694 discrete samples (1- 3 sec) spanning 30 minutes duration were labeled manually. The details of the manual label data are given in Table \ref{tab:ml}. We faced some challenges in several cases to recognize the honk of medium-weight vehicles (MWV), as shown in Table \ref{tab:ml}. For example, the MWV-honking is often labeled as HWV-honking by the personnel. Similarly, the honk of MWV is tagged as LWV-honk in many situations. Results show that our manual labeling process gives a total dissimilarity of around 7\%. This dissimilarity might occur due to several issues: \textit{1)} raw audio data includes a variety of ambient noises, \textit{2)} honk sounds came from different sides of the lane, which is not captured in the videos and the signature of different vehicle honks are sometimes confusing due to their similar pattern, and \textit{3)} occasionally, multiple vehicular honks may overlap in time. To improve the accuracy of our training process, we remove mismatched data from our contemplation. However, these remaining samples are too few for training a model, which will result in more errors and low accuracy. Therefore, we require an automated data labeling model to annotate our data with higher accuracy.

Autoencoder(AE) and semi-supervised generative adversarial network (SGAN) based labeled data set preparation has been addressed in many works. The AE-based multi-class data labeling techniques are discussed in \cite{bank2020autoencoders,wicker2016nonlinear,law2019multi,aamir2021deep}. Authors \cite{wicker2016nonlinear} introduced a multi-class data labeling autoencoder model, where they have compressed the labels using AE by eliminating the non-linear dependencies. A stacked autoencoder network (SAE) is developed in \cite{law2019multi} to produce a discriminating and decreased input representation of the multi-label data. Similarly, the contractive autoencoder (CAE) model uses layered architecture followed by a feed-forward mechanism to encode unlabeled training data \cite{aamir2021deep}. Apart from these studies, the researchers also developed SGAN-based data labeling models \cite{khan2022semi,amin2020semi}. \cite{amin2020semi} used both SGAN and transfer learning models for labeling the data and model training at a time. In our previous work \cite{maity2022dehonk}, we considered the SGAN model to label the data into two classes (honk or non-honk) to identify vehicular honks from traffic data in presence of ambient noise. Researchers also tried to combine the autoencoder and adversarial learning for fault diagnosis in \cite{wen2023novel}.

\begin{figure}
    \centering
    \includegraphics[width=7cm, height= 5cm]{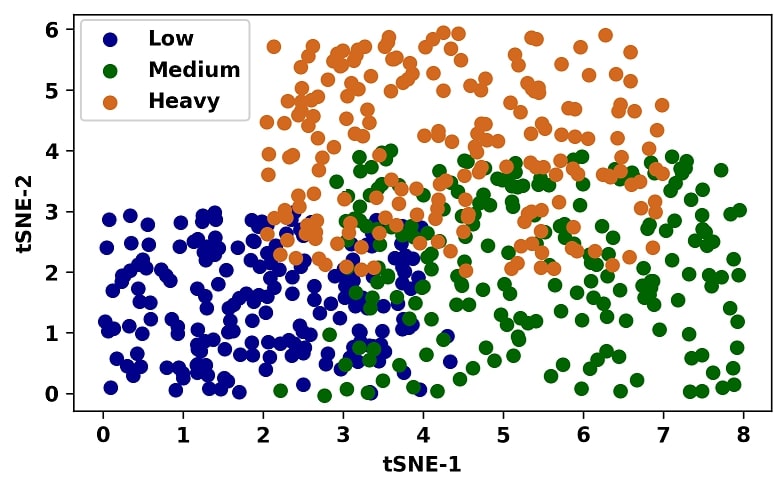}
    \caption{Overlapped data point, focusing the signature of the MWV sometimes belongs to either HWV or LWV}
    \label{fig:tsne}
\end{figure}

Inspired by this, at first, we tried to label our collected data using the existing autoencoder model, SGAN, SAE \cite{law2019multi}, CAE \cite{aamir2021deep}. Although these techniques give satisfactory outcomes in earlier cited works, they failed to obtain decent accuracy in our work due to the above-mentioned issues. Fig. \ref{fig:tsne} shows the t-SNE plot of the data samples. To improve the clarity of the plot, we have chosen 30\% of the actual data set and noticed that, only in a few cases, the honking signature of LWV and HWV overlapped, but in most of the cases, MWV honking pattern is confusing due to similarities between patterns, either HWV or LWV. Therefore, we need a model that can distinguish all the vehicle types efficiently. In this paper, we have extended the autoencoder (AE) model and proposed two data labeling models, i.e.,  Multi-label Autoencoder (MAE) and Multi-label Autoencoder Generative Adversarial Network (MAEGAN). 

\begin{table*}[]
\centering
\caption{Distribution of manual label data}
\begin{tabular}{|lllll|}
\hline
\rowcolor[HTML]{C0C0C0} 
\multicolumn{5}{|c|}{\cellcolor[HTML]{C0C0C0}\textit{Personnel 1},  \underline{Personnel 2}, \textbf{Personnel 1 $\cap$ Personnel 2 }}                                                                                                                                                  \\ \hline
\rowcolor[HTML]{EFEFEF} 
\multicolumn{1}{|l|}{\cellcolor[HTML]{EFEFEF}} & \multicolumn{1}{l|}{\cellcolor[HTML]{EFEFEF}Non-honk} & \multicolumn{1}{l|}{\cellcolor[HTML]{EFEFEF}LWV} & \multicolumn{1}{l|}{\cellcolor[HTML]{EFEFEF}MWV} & HWV     \\ \hline
\multicolumn{1}{|l|}{Non-honk}                 & \multicolumn{1}{l|}{\textit{740},\underline{738},\textbf{735}}                          & \multicolumn{1}{l|}{\textit{11},\underline{12},\textbf{9}}                       & \multicolumn{1}{l|}{\textit{6},\underline{7},\textbf{6}}                         & \textit{3},\underline{2},\textbf{2}     \\ \hline
\multicolumn{1}{|l|}{LWV}                      & \multicolumn{1}{l|}{\textit{1},\underline{2},\textbf{1}}                              & \multicolumn{1}{l|}{\textit{437},\underline{439},\textbf{434}}                     & \multicolumn{1}{l|}{\textit{10},\underline{8},\textbf{7}}                        & \textit{2},\underline{1},\textbf{1}     \\ \hline
\multicolumn{1}{|l|}{MWV}                      & \multicolumn{1}{l|}{\textit{1},\underline{0},\textbf{0}}                              & \multicolumn{1}{l|}{\textit{11},\underline{10},\textbf{8}}                       & \multicolumn{1}{l|}{\textit{318},\underline{321},\textbf{311}}                     & \textit{10},\underline{9},\textbf{7}    \\ \hline
\multicolumn{1}{|l|}{HWV}                      & \multicolumn{1}{l|}{\textit{0},\underline{0},\textbf{0}}                              & \multicolumn{1}{l|}{\textit{1},\underline{1},\textbf{1}}                         & \multicolumn{1}{l|}{\textit{5},\underline{6},\textbf{4}}                        & \textit{206},\underline{204},\textbf{203} \\ \hline
\end{tabular}
\label{tab:ml}
\end{table*}

\begin{figure*}[h]
    \centering
    \includegraphics[width=\linewidth]{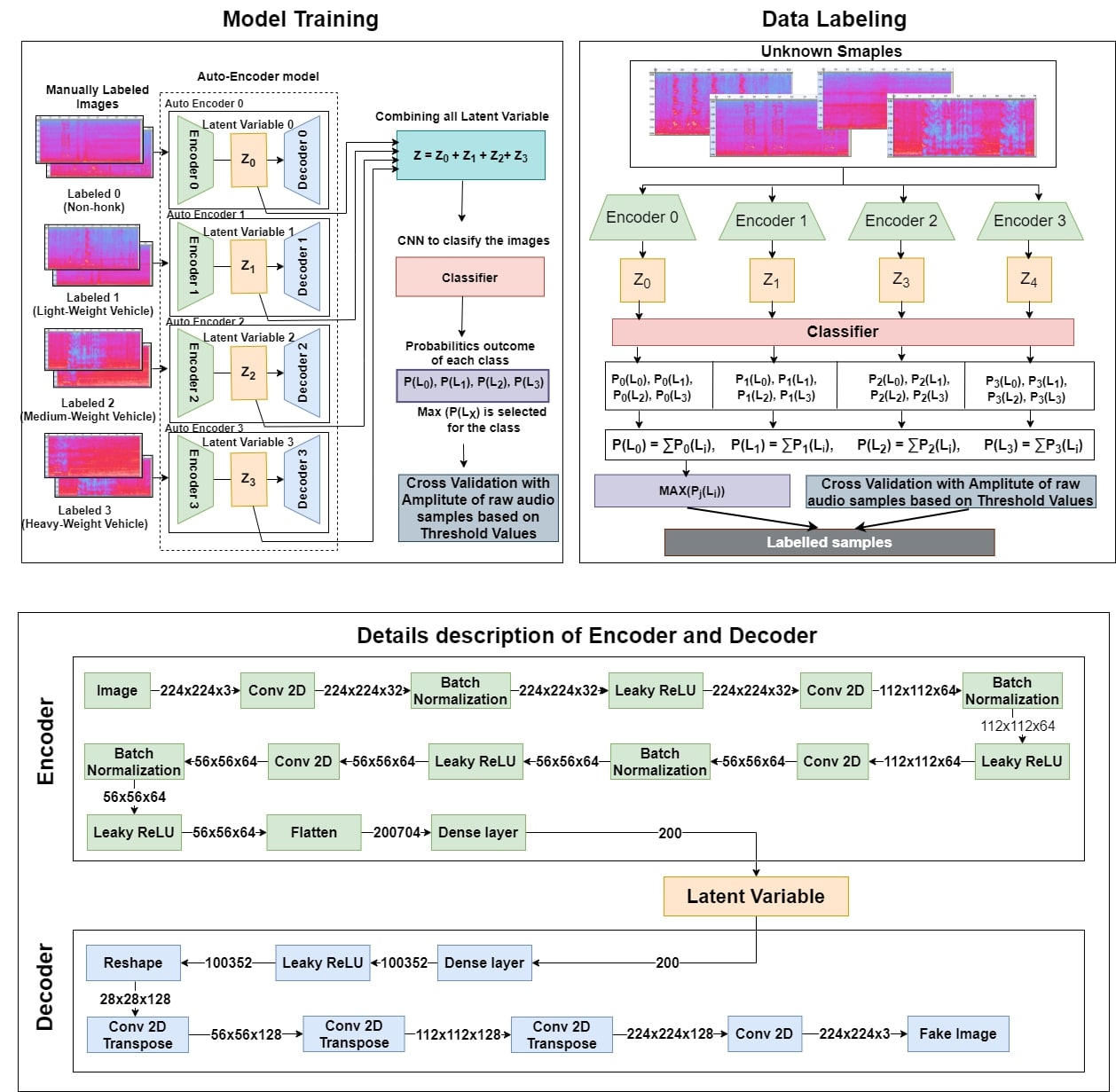}
    \caption{The architecture of our proposed Multi-label Auto Encoder model (MAE)}
    \label{fig:cae}
\end{figure*}

\textit{\textbf{(i) Proposed Multi-label Autoencoder (MAE):}}  Autoencoder is the combination of encoder (squeeze the input into a lower-dimensional code) and decoder (reconstructs the image from the lower-dimensional image) along with the latent space (compressed input). In this work, we have used four different autoencoders (AE0 to AE3), where AE0 is used for the images of labeled 0, AE1 is used for the images of labeled 1, and so on. For all four encoders, four latent space(Z0, Z1, Z2, Z3) is generated. Later we combine all the latent space and create a single latent space Z. Now, we pass this Z to CNN models for the vehicular honk classification. We calculated the probabilities outcome of each class, and the highest probability is considered as a selected class. Moreover, we cross-validate the selected class with the amplitude value of the respective raw audio samples. In this way, we have trained our MAE model for data labeling. Once the training is done, for data labeling, we take the unlabeled data and pass it to the trained encoder, from which latent space is generated for all the classes, and then pass it to the CNN model and select the highest probabilistic class with the cross-validation with corresponding amplitude values. The architecture of the proposed model is illustrated in Fig. \ref{fig:cae}.

For our proposed MAE, Latent space Z is represented as 
\begin{equation}
    Z_i = \sigma (W_i x_i + b_i)
\end{equation}  
where $x_i$ is the original input image, $b_i$ is the bias, $W_i$ is the weight and $\sigma$ is the activation function. Similarly, decoding is represented as 
\begin{equation} 
\widehat{x_i} = \widehat{\sigma} (\widehat{W}z_i + \widehat{b_i})
\end{equation}
Loss function is formulated as 
\begin{equation} \mathcal{L}_i = \frac{1}{N_i} \sum_{n=1}^{N_i} (x_i - \widehat{x_i})^2
\end{equation} 
In all cases, the value of $i$ ranges between 0 to 3.

\textit{\textbf{(ii) Proposed Multi-label Autoencoder Generative Adversarial Network (MAEGAN):}} 
A generative adversarial network (GAN) has two components. One is a generator, which generates the fake images, and another one is a discriminator, which distinguishes the real and fake images. In MAEGAN, instead of one generator and one discriminator, we have used four generators (G0-G3) and four discriminators (D0-D3) to label four different classes, i.e., non-honk, LWV, MWV, HWV. In our work, G0 generates fake images for labeled 0, and D0 is used to distinguish the real and fake images of labeled 0. The remaining generators and discriminators are used for other classes of images. The proposed MAEGAN model uses the trained latent space of the MAE model as a generator. The intuition behind using the latent space of the MAE model is that it is already trained with all the different labels of the image classes. Hence, the generator will generate more accurate images, which may be hard to classify by the discriminator. Similar to the proposed MAE model, we have used the same classifier to classify different classes of honks. It is found that the proposed MAE performs better than MAEGAN, which is shown in the result section in detail.

\subsection{Data augmentation and class balancing} \label{sec:dac}
By combining manual and MAE model generated samples, we obtained a total of 56399 samples, where 26\%, 20\%, and 12\% of data belongs to LWV, MWV, and HWV classes, respectively, and the remaining 42\% of data is in non-honk class. Due to the disparity in size among honk classes, we faced an imbalanced classification problem in our work. 
Class balancing in such a scenario may lead to losing a large amount of data, which may further result in the reduction of training samples. To overcome this problem, we do the data augmentation on the spectrogram image rather than raw audio. There exist multiple methods of data augmentation (like time warping, time masking, frequency masking, etc.) to enhance the data volume artificially. 

\begin{figure*}[h]
\centering
\subfigure[]{\label{fig:mask1}{\includegraphics[width=7cm, height= 4.5cm]{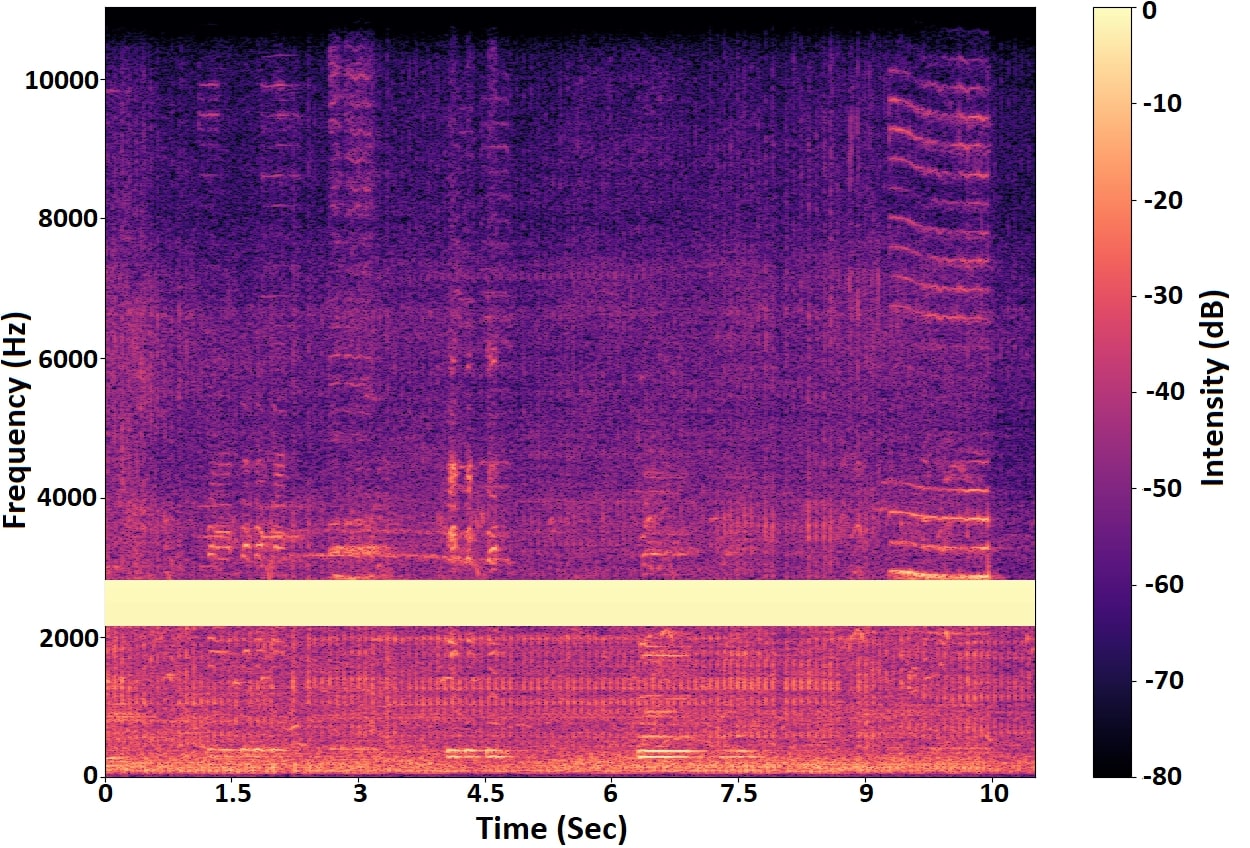}}}
\subfigure[]{\label{fig:mask2}{\includegraphics[width=7cm, height= 4.5cm]{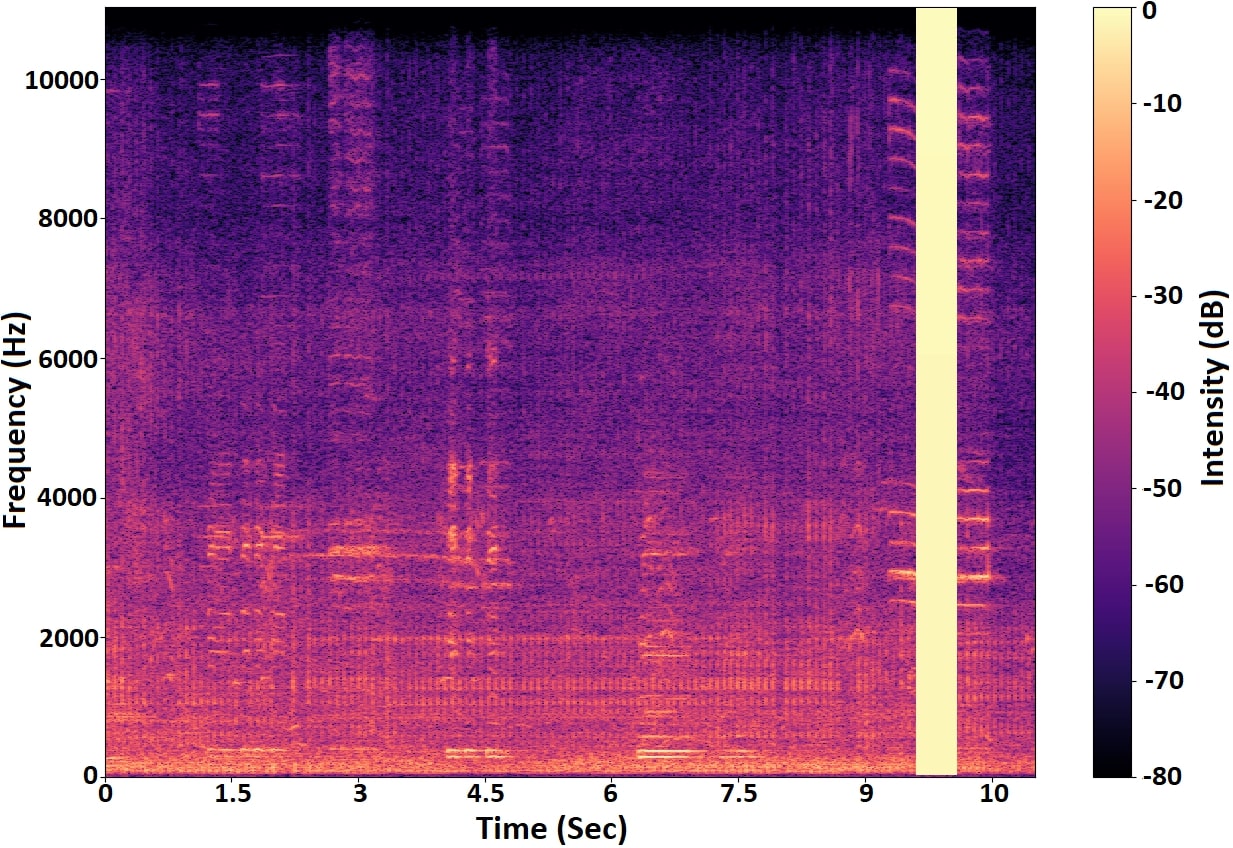}}}
\caption{Illustration of data augmentation on a spectrogram image with (a) Time masking, (b) Frequency masking }
\label{fig:mask}
\end{figure*}

In our work, we have used time and frequency mask techniques to augment the volume of data. In these techniques, some vertical bars are randomly placed to conceal the time masks and horizontal bars to conceal frequency masks on the spectrogram images, respectively. To visualize this, we have considered the spectrogram image shown in Fig. \ref{fig:spe1}. We see the spectrogram image with vertical and horizontal bars showing time mask and frequency mask-based augmentation in Fig. \ref{fig:mask1} and Fig. \ref{fig:mask2}, respectively. After performing data augmentation, the data volume is increased by around 60\%. Next, we do the class balancing, which contains a total of 22 hours of data, and split the dataset for training and testing. The details of the distribution are shown in Table \ref{tab:dd}.

\begin{table*}[h]\footnotesize
\centering
\caption{Distribution of data before and after class balancing \cite{maity2022dehonk}}
\begin{tabular}{|c|c|c|c|c|cc|}
\hline
\rowcolor[HTML]{C0C0C0} 
         & Manual Label Data & Labeled by MAE & After Augmentation & {\cellcolor[HTML]{C0C0C0}\begin{tabular}[c]{@{}c@{}}After Class\\ Balancing\end{tabular}} & \multicolumn{2}{c|}{\cellcolor[HTML]{C0C0C0}\begin{tabular}[c]{@{}c@{}}Distribution of data for\\ training and testing\end{tabular}} \\ \hline
\rowcolor[HTML]{EFEFEF} 
Class    & \#samples    & \#samples & \#samples     & \#samples  & \multicolumn{1}{c|}{\cellcolor[HTML]{EFEFEF}Training set}                                & Testing set                               \\ \hline
Total    & 1694              & 54705          & 134286             & 79608           & \multicolumn{1}{c|}{63684}                                                               & 15924                                     \\ \hline
Non-honk & 741               & 22853          & 68559              & 19902           & \multicolumn{1}{c|}{15921}                                                               & 3981                                      \\ \hline
LWV      & 440               & 14170          & 42511              & 19902           & \multicolumn{1}{c|}{15921}                                                               & 3981                                      \\ \hline
MWV      & 320               & 11048          & 33146              & 19902           & \multicolumn{1}{c|}{15921}                                                               & 3981                                      \\ \hline
HWV      & 193               & 6634           & 19902              & 19902           & \multicolumn{1}{c|}{15921}                                                               & 3981                                      \\ \hline
\end{tabular}
\label{tab:dd}
\end{table*}

\subsection{Choice of classification models} \label{sec:ms}
In several domains, such as object detection, image recognition, and image classification, Convolutional Neural Networks (CNNs) are widely utilized \cite{piczak2015environmental,salamon2017deep}. Transfer learning-based CNN models \cite{demir2020new} are often employed to achieve higher accuracy. These models leverage pre-training on the ImageNet dataset, which consists of over 14 million images and is accessible through trusted public Keras libraries. By utilizing pre-trained models, training time is reduced, and there is less dependability on having a large training dataset.

\textbf{\textit{(i) Pre-trained CNN models:}} For classification, we have chosen four pre-trained CNN models, namely MobileNet, ShuffleNet, ResNet 50, and Inception V3. These models were selected by considering two environments: (i) low-end systems and (ii) hardware accelerator systems. We have used two lightweight models for low-end systems: MobileNet and ShuffleNet. On the other hand, ResNet 50 and Inception V3 are used as Heavyweight models for hardware accelerator systems. For our problem, these models have been trained on our dataset. Additionally, pre-trained models are fine-tuned by adding a dense layer to overcome the problem of overfitting. 

\begin{figure*}[h]
    \centering
    \includegraphics[width=\linewidth]{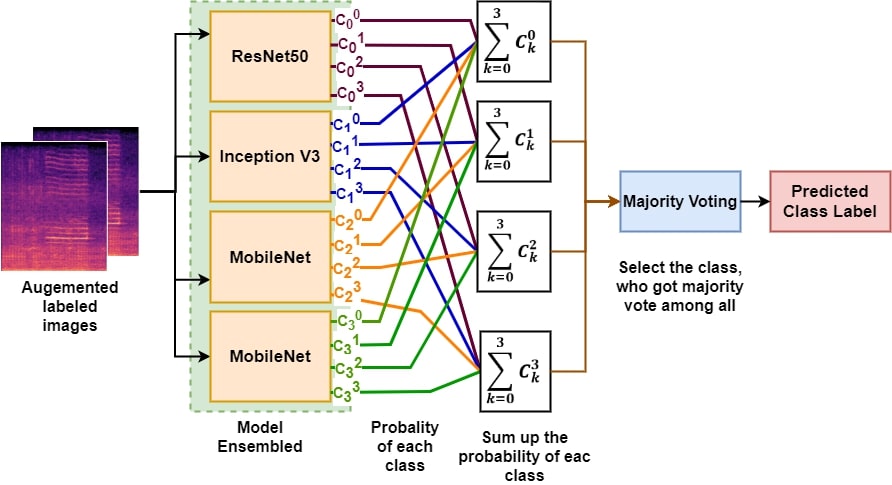}
    \caption{Architecture of proposed EnTL for honk classification}
    \label{fig:entl}
\end{figure*}

\textbf{\textit{(ii) Proposed EnTL model: }}
The ensemble model leverages the idea that aggregating the predictions of multiple models can often lead to more accurate and robust predictions than using a single model. To improve the classification accuracy, we propose a transfer learning-based ensemble method, i.e., Ensambled Transfer Learning (EnTL), combining four fine-tuned pre-trained models (i.e., MobileNet, ShuffleNet, ResNet 50, Inception V3). All the models are trained independently on different subsets of the training data, created through random sampling with replacement. Each model provides a probabilistic score for each class using the 'softmax' function. For an input image $x$, a model can generate four probability scores, one for each class. Let $C_{k}^{i}$ be the probability score of model $k$ on class $i$, where $k$ = 0,1,2, ... (M-1); $M$ represents the number of models and $i$ ranges from 0 to 3. We sum up the probability of each class $i$ generated by $M$ models, which is a 1-D vector for $x$. From this vector, we select the class having the highest value using majority voting as a final class for the input image sample $x$ (see in Equation \ref{eq:e4}). Fig. \ref{fig:entl} illustrates the architecture of the proposed model(EnTL).

\begin{equation}
    Final_{outcome} = \underset{i}{\arg\max} \sum_{k=0}^{M-1}C_{k}^{i}(x) \label{eq:e4}
\end{equation}

\section{Experimental result analysis} \label{sec:ra}
The performance of all the models for honk classification is evaluated using accuracy, Matthews correlation coefficient (MCC), F1 score, precision, recall, kappa statistics, and area under ROC curve (ROC AUC). In Section \ref{sec:pdl} and \ref{sec:phc}, we have compared the performance of different data labeling and classification models, respectively. 
To show the superiority of our proposed model, we have compared our proposed classification model with baselines in Section \ref{sec:cbw}. In Section \ref{sec:ci}, we have inferred context considering real-life scenarios, and at the end, we have discussed the overall importance of our work in Section \ref{sec:dr}.  

\subsection{Comparison with existing data labeling models} \label{sec:pdl}
\begin{figure}
    \centering
    \includegraphics[width=7cm, height= 5cm]{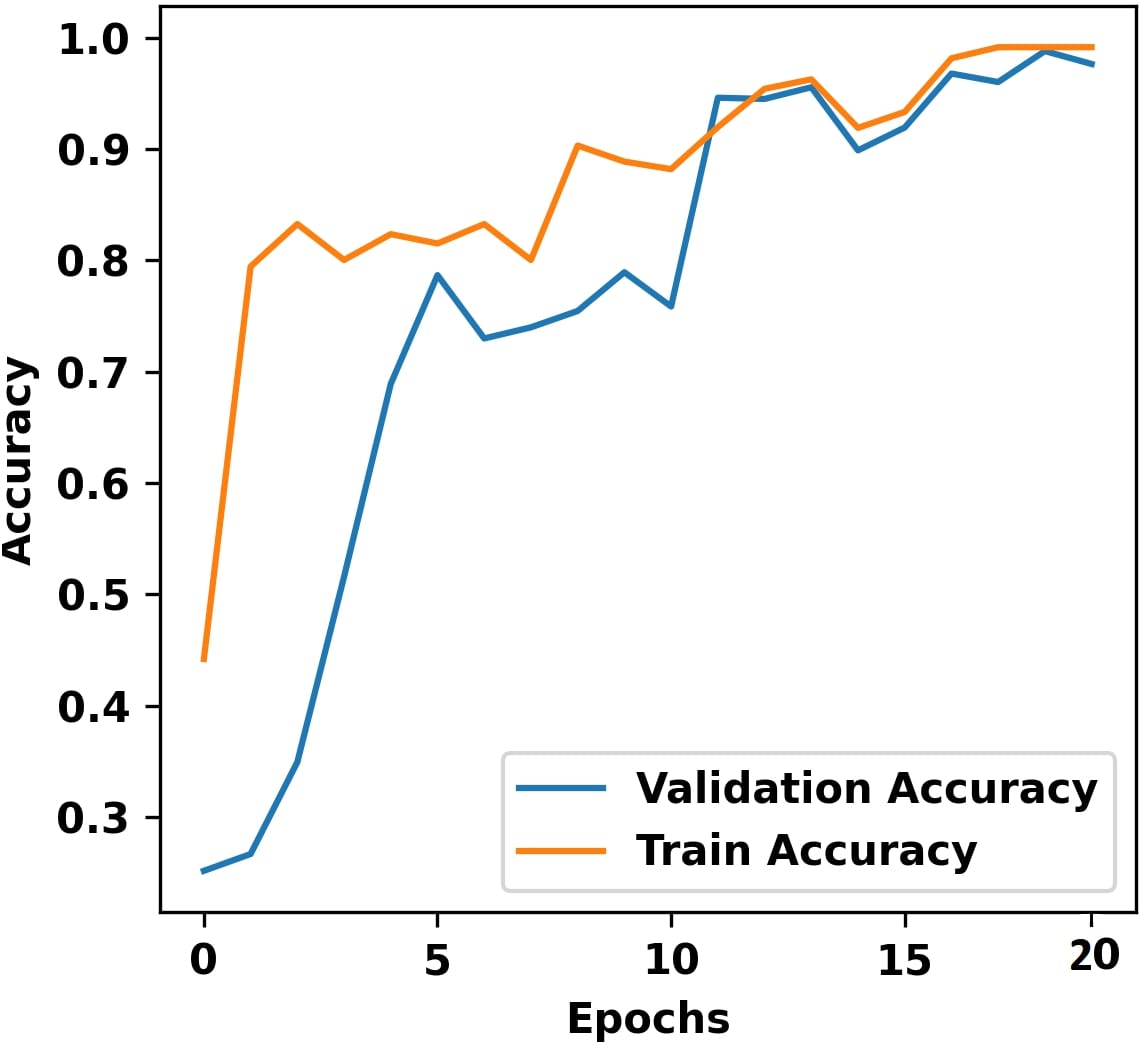}
    \caption{Epoch vs. accuracy plot at the time of the training of MAE model}
    \label{fig:CATT}
\end{figure}

To evaluate the effectiveness of the proposed data labeling models i.e., MAE and MAEGAN, we perform a comparative analysis against the conventional AE model, GAN, CAE model \cite{aamir2021deep}, and SAE model. All these models are discussed previously in Section \ref{sec:dl}). Table \ref{tab:dala} summarizes the experimental results of all models, MAE and MAEGAN. It can be seen that the CAE model performs better in contrast to existing deep learning models i.e., the conventional AE, GAN, and SAE models. But the results indicate that our proposed MAE and MAEGAN models outperform the CAE model. In fact, the performance of MAE and MAEGAN models is $\sim$18\% better than the CAE model. This is because the existing algorithms failed to distinguish the honking pattern of MWV and HWV in several cases due to similarities between the data samples, as shown in Fig. \ref{fig:tsne}. As a result, their accuracy dropped. But our multi-label autoencoder models clearly distinguish the honking pattern of MWV, HWV, LWV, and even non-honks and thus give higher accuracy. Furthermore, comparing the performance of the MAE model with the MAEGAN model, we see that the MAE shows $\sim$2.5\% better outcome than the MAEGAN. Due to this performance superiority, we have used the MAE model for labeling the unlabeled data in our work. To verify the correctness of the proposed MAE, we have performed a ground truth verification discussed in the following sub-section. A total of 1694 samples are labeled manually, and the remaining data are labeled using MAE. The entire data set is divided into five distinct groups, and we fed each group of data one by one to the MAE model for labeling. A total of 54705 samples are labeled by the proposed MAE model with more than 97 \% accuracy. Epoch-wise training and validation accuracy are also depicted in Fig. \ref{fig:CATT}.


\begin{table*}[h]
\tiny
\centering
\caption{Performance analysis of different models for data labeling}
\begin{tabular}{|l|cccc|cc|}
\hline
\rowcolor[HTML]{C0C0C0} 
\multicolumn{1}{|c|}{\cellcolor[HTML]{C0C0C0}}                                             & \multicolumn{4}{c|}{\cellcolor[HTML]{C0C0C0}\textbf{Baseline Models}}                                                                                                                                                                                                                  & \multicolumn{2}{c|}{\cellcolor[HTML]{C0C0C0}\textbf{Proposed Models}}       \\ 
\rowcolor[HTML]{C0C0C0} 
\multicolumn{1}{|c|}{\multirow{-2}{*}{\cellcolor[HTML]{C0C0C0}\begin{tabular}[c]{@{}c@{}}\textbf{Evaluation}\\ \textbf{Metric}\end{tabular}}} & \multicolumn{1}{c|}{\begin{tabular}[c]{@{}c@{}}\textbf{AE}\\ \cite{bank2020autoencoders}\end{tabular}} & \multicolumn{1}{c|}{\begin{tabular}[c]{@{}c@{}}\textbf{SAE}\\ \cite{law2019multi}\end{tabular}} & \multicolumn{1}{c|}{\begin{tabular}[c]{@{}c@{}}\textbf{CAE}\\ \cite{aamir2021deep}\end{tabular}} & \multicolumn{1}{c|}{\begin{tabular}[c]{@{}c@{}}\textbf{SGAN}\\ \cite{maity2022dehonk}\end{tabular}} & \multicolumn{1}{c|}{\cellcolor[HTML]{C0C0C0}\textbf{MAE}} & \textbf{MAEGAN} \\ \hline
Accuracy                                                                                   & \multicolumn{1}{c|}{61.65\%}                                                         & \multicolumn{1}{c|}{70.08\%}                                                  & \multicolumn{1}{c|}{79.48\%}                                                      & 66.19\%                              & \multicolumn{1}{c|}{\textbf{97.64\%}}                     & 94.99\%         \\ \hline
MCC                                                                                        & \multicolumn{1}{c|}{0.43}                                                            & \multicolumn{1}{c|}{0.62}                                                     & \multicolumn{1}{c|}{0.71}                                                      & 0.49                                 & \multicolumn{1}{c|}{\textbf{0.97}}                        & 0.93            \\ \hline
F1 Score                                                                                   & \multicolumn{1}{c|}{0.54}                                                            & \multicolumn{1}{c|}{0.48}                                                     & \multicolumn{1}{c|}{0.71}                                                      & 0.59                                 & \multicolumn{1}{c|}{\textbf{0.96}}                        & 0.92            \\ \hline
Precision                                                                                  & \multicolumn{1}{c|}{0.61}                                                            & \multicolumn{1}{c|}{0.83}                                                     & \multicolumn{1}{c|}{0.86}                                                      & 0.65                                 & \multicolumn{1}{c|}{\textbf{0.96}}                        & 0.93            \\ \hline
Recall                                                                                     & \multicolumn{1}{c|}{0.52}                                                            & \multicolumn{1}{c|}{0.52}                                                     & \multicolumn{1}{c|}{0.69}                                                      & 0.57                                 & \multicolumn{1}{c|}{\textbf{0.97}}                        & 0.93            \\ \hline
Kappa Statistics                                                                           & \multicolumn{1}{c|}{0.41}                                                            & \multicolumn{1}{c|}{0.56}                                                     & \multicolumn{1}{c|}{0.69}                                                      & 0.49                                 & \multicolumn{1}{c|}{\textbf{0.97}}                        & 0.93            \\ \hline
Roc AUC                                                                                    & \multicolumn{1}{c|}{0.69}                                                            & \multicolumn{1}{c|}{0.71}                                                     & \multicolumn{1}{c|}{0.80}                                                      & 0.74                                 & \multicolumn{1}{c|}{\textbf{0.98}}                        & 0.96            \\ \hline
\end{tabular}
\label{tab:dala}
\end{table*}

\subsection {Ground truth verification of labeled data w.r.t data models}
\begin{figure*}
    \centering
    \includegraphics[width=16cm, height= 12cm]{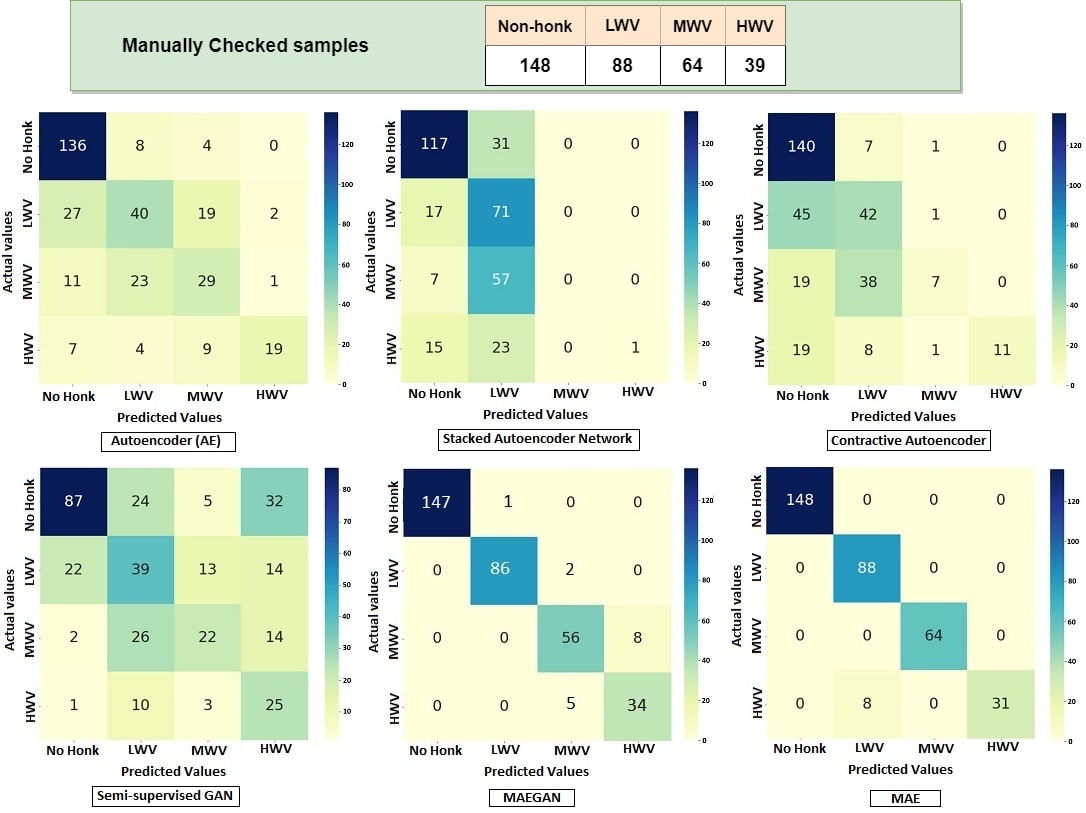}
    \caption{Proof of correctness of MAE model by comparing existing models data and manual data (we test it by considering 5\% of sample data, selected randomly)}
    \label{fig:ldgt}
\end{figure*}

To check the correctness of our proposed MAE model, we picked up 340 samples randomly for validation and labeled them using MAE and MAEGAN models. The samples were also fed to other existing data models (AE, GAN, SAE, CAE) for labeling and then compared with MAE and MAEGAN models. For ground truth verification, these 340 samples are further labeled manually by two volunteers, where the number of LWV, MWV, and HWV honking samples are 88, 64, and  39, respectively. The results are shown in Fig. \ref{fig:ldgt}. We see that only one honk of MWV is misclassified as the honk of HWV, and 8 HWV samples are detected as LWV honks by the MAE model. It implies that out of 340 samples, only 9 samples are wrongly tagged by the proposed MAE model, while the misclassified samples generated by other models are 16, 115, 150, 139, and 166 for MAEGAN, AE, SAE, CAE, SGAN models respectively. The result signifies that our proposed MAE model can correctly label the unlabeled samples.

\subsection{Classification performance evaluation of pre-trained models and EnTL model} \label{sec:phc}
Table \ref{tab:phc} illustrates the performance comparison for all the pre-trained models and EnTL for classifying honks using the chosen performance measures. Here, we have used 79608 labeled samples, where 63684 samples were for training and 15924 samples were for testing. We have trained all pre-trained models using the Adam optimizer for 20 epochs. The input image size is 224 × 224 × 3, the learning rate is set at 0.001, and the batch size is set to 312. Results indicate that lightweight models i.e., MobileNet and ShuffleNet, have achieved test accuracy at around 88.82\% and 78.59\%, respectively. 

\begin{figure*}
\centering
\subfigure[]{\label{fig:c1}{\includegraphics[width=7cm, height= 5cm]{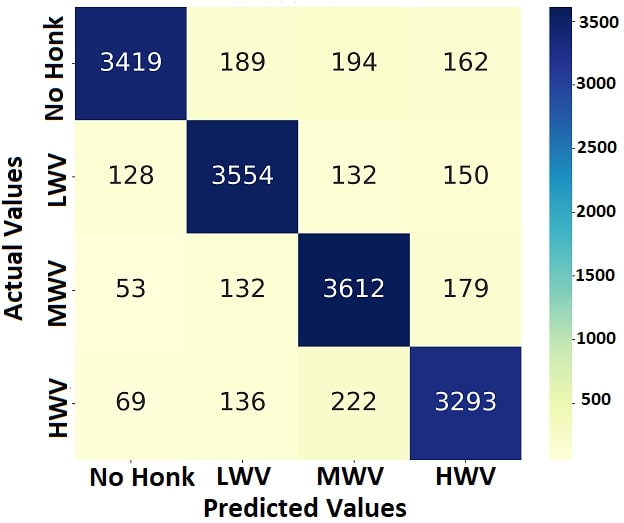}}}
\subfigure[]{\label{fig:c2}{\includegraphics[width=7cm, height= 5cm]{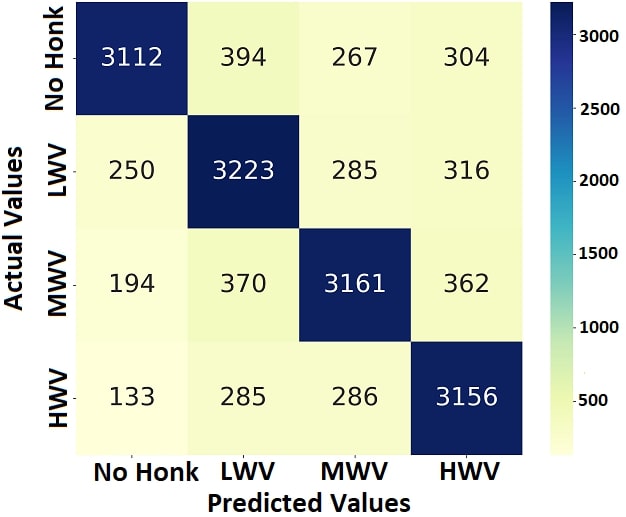}}}
\subfigure[]{\label{fig:c3}{\includegraphics[width=7cm, height= 5cm]{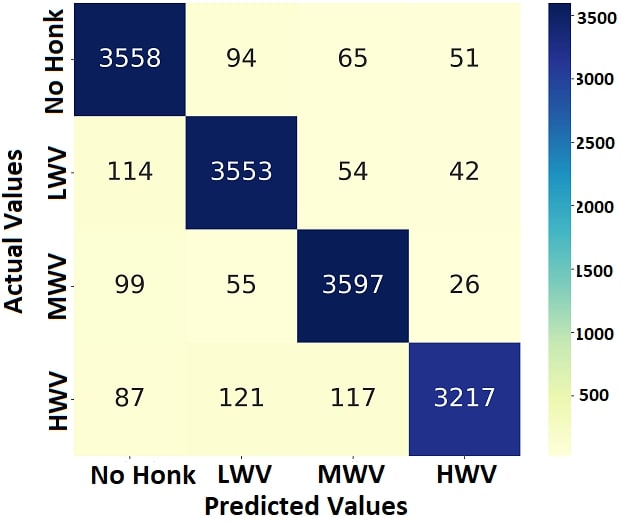}}}
\subfigure[]{\label{fig:c4}{\includegraphics[width=7cm, height= 5cm]{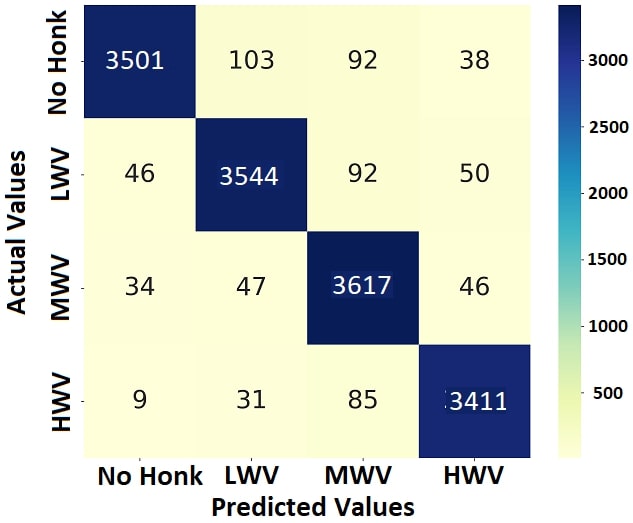}}}
\subfigure[]{\label{fig:c5}{\includegraphics[width=7cm, height= 5cm]{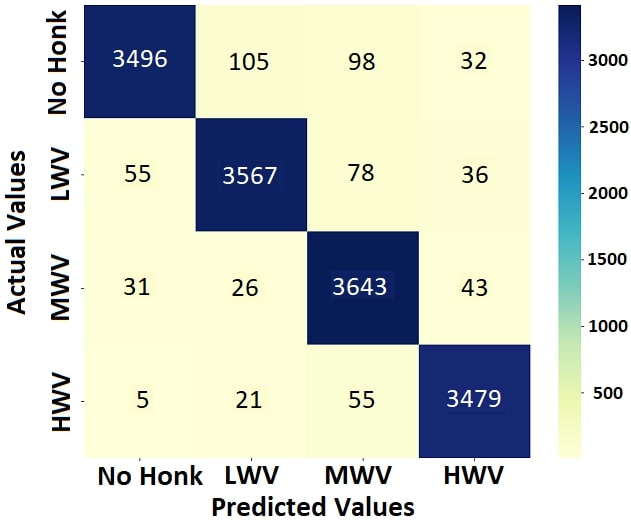}}}
\caption{Confusion matrix of (a) MobileNet (b) ShuffleNet (c) ResNet50 (d) Inception V3, and (e) EnTL}
\label{fig:con}
\end{figure*}

On the other hand, ResNet 50 and Inception V3 have achieved accuracies of 93.77\%  and 95.17\% , respectively. among the pre-trained models, Inception V3 gives better results in our dataset. Nevertheless, the proposed EnTL model surpasses that of Inception V3 with an accuracy of 96.72\% and demonstrates classification accuracy of 95.47\%, 97.32\%, and 97.72\% for LWV, MWV, and HWV, respectively. Additionally, the attained values of MCC, F1-score, precision, recall, Kappa statistics, and ROC-AUC demonstrate that EnTL outperforms other models. The confusion matrix, shown in Fig. \ref{fig:con}, is calculated for each model. We have found that the misclassification (false positive) rate is significantly less in EnTL. Only a few non-honk samples (2.7\%) are detected as LWV because sometimes the pattern of engine noise/several other noises becomes similar to the LWV. Similarly, 2.41\% samples of LWV are classified as MWV due to the same type of honking pattern that may arise in the case of motorbikes and four-wheeler vehicles. The remaining misclassification rate is less than 1\%, which is negligible. As a note, the classification performance of the proposed EnTL is better than any single transfer learning model and can be used as a model to classify vehicular honks even in the presence of ambient noises. 


\begin{table*}
\caption{Performance analysis of different models for honk classification}
\begin{tabular}{|l|cc|cc|c|}
\hline
\rowcolor[HTML]{C0C0C0} 
\multicolumn{1}{|c|}{\cellcolor[HTML]{C0C0C0}}                                             & \multicolumn{2}{c|}{\cellcolor[HTML]{C0C0C0}\textbf{Light weight model}}              & \multicolumn{2}{c|}{\cellcolor[HTML]{C0C0C0}\textbf{Heavy weight model}}  & \multicolumn{1}{c|}{\cellcolor[HTML]{C0C0C0}\textbf{Proposed model}}              \\  
\rowcolor[HTML]{C0C0C0} 
\multicolumn{1}{|c|}{\multirow{-2}{*}{\cellcolor[HTML]{C0C0C0}\textbf{Evaluation Metric}}} & \multicolumn{1}{c|}{\cellcolor[HTML]{C0C0C0}\textbf{MobileNet}} & \textbf{ShuffleNet} & \multicolumn{1}{c|}{\cellcolor[HTML]{C0C0C0}\textbf{ResNet 50}} & \textbf{Inception V3} & \multicolumn{1}{c|}{\cellcolor[HTML]{C0C0C0}\textbf{EnTL}}\\ \hline
Accuracy                                                                                   & \multicolumn{1}{c|}{88.82\%}                                    & 78.59\%             & \multicolumn{1}{c|}{93.77\%}                                    & 95.17\% & \textbf{96.72}\%   \\ \hline
MCC                                                                                        & \multicolumn{1}{c|}{0.85}                                       & 0.72                & \multicolumn{1}{c|}{0.92}                                       & 0.94 & \textbf{0.96 }       \\ \hline
F1 Score                                                                                   & \multicolumn{1}{c|}{0.89}                                       & 0.79                & \multicolumn{1}{c|}{0.94}                                       & 0.95 & \textbf{0.97}         \\ \hline
Precision                                                                                  & \multicolumn{1}{c|}{0.89}                                       & 0.79                & \multicolumn{1}{c|}{0.94}                                       & 0.95 & \textbf{0.97}         \\ \hline
Recall                                                                                     & \multicolumn{1}{c|}{0.89}                                       & 0.79                & \multicolumn{1}{c|}{0.94}                                       & 0.95 & \textbf{0.97}         \\ \hline
Kappa Statistics                                                                           & \multicolumn{1}{c|}{0.85}                                       & 0.71                & \multicolumn{1}{c|}{0.92}                                       & 0.94 & \textbf{0.96 }        \\ \hline
ROC AUC                                                                                    & \multicolumn{1}{c|}{0.93}                                       & 0.86                & \multicolumn{1}{c|}{0.96}                                       & 0.97& \textbf{0.98 }        \\ \hline
\end{tabular}
\label{tab:phc}
\end{table*}

\subsection{Significance of dataset} \label{sec:id}
\begin{figure}
    \centering
    \includegraphics[width=8cm, height= 6.3cm]{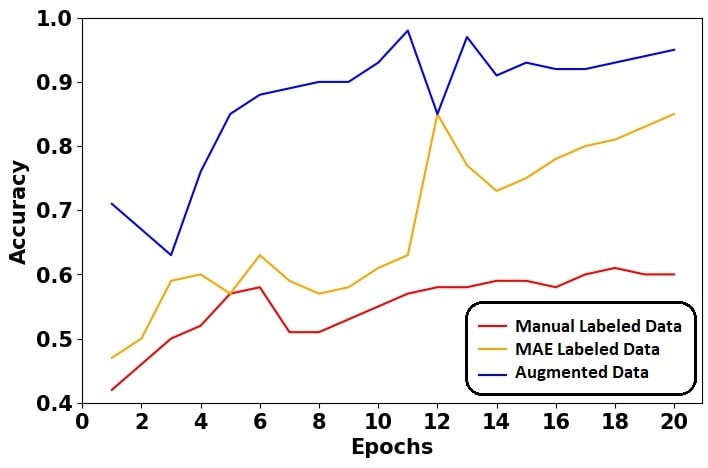}
    \caption{Training accuracy w.r.t different level of dataset}
    \label{fig:id}
\end{figure}

To show the importance of our curated dataset, we have trained the best-performing model EnTL for different datasets and plotted the training accuracy in Fig. \ref{fig:id}. At first, we trained the model with manually labeled samples. The training accuracy dropped since our manually labeled samples(1694) were few. Next, we trained the model using MAE-generated labeled samples (54705) and obtained better accuracy than training with manual label samples. Finally, the model was trained with the enriched dataset(79806) generated by data augmentation techniques, and we achieved a very good training accuracy. Hence, our proposed methodology and the dataset play a crucial role in this study as well as the annotate dataset can be useful for other research.  

\subsection{Performance comparison of EnTL with the recent literature of honk classification} \label{sec:cbw}
In this section, the performance of the EnTL model is compared with baseline work using chosen performance metrics. Here, we have considered SBCNN \cite{salamon2017deep}, Dilated CNN \cite{chen2019environmental}, CNN \cite{demir2020new}, and TFCNN \cite{mu2021environmental} as baseline works. The configuration settings of each baseline model are represented in Table \ref{tab:blc}. The detailed results are presented in Table \ref{tab:blc1}, from where we notice that the performance of EnTL is increased by at least $\sim$9\% and at most $\sim$21\% in contrast to other models. The baseline models reported higher sound classification accuracy in publicly available datasets like ESC-10, ESC-50, or UrbanSound8k, where ambient noise was either absent or not considered. Moreover, each audio file's duration is much less in these datasets. In contrast, our data was collected from a real environment in the presence of various ambient noises. Hence, the accuracy fell sharply when we fed our dataset to the existing models. However, the Dilated CNN provides the best output among all the baseline models. When we compare our model with Dilated CNN model, the performance of our model is increased by 10.15\%, 17.07\%, 11.49\%, 10.22\%, 12.79\%, 17.07\%, 7.69\% for accuracy, MCC, F1-score, Precision, Recall, Kappa Statistics, ROC AUC. The EnTL model is good for classifying honk based on the vehicle types in the presence of ambient noises.

\begin{table*}[]
\footnotesize
\caption{Specifications of the baseline models \cite{maity2022dehonk}}
\centering
\begin{tabular}{|c|c|c|c|c|c|c|}
\hline
\rowcolor[HTML]{C0C0C0} 
Models  & Conv2d & Max polling & Up Sampling & Dense Layer & \begin{tabular}[c]{@{}c@{}}Batch \\ Normalization\end{tabular} & \begin{tabular}[c]{@{}c@{}}Activation\\  Function\end{tabular} \\ \hline
SB-CNN \cite{salamon2017deep}  & 3      & 2           & -           & 3           & 3                                                              & Softmax                                                        \\ \hline
Dilated CNN \cite{chen2019environmental} &5 & 1 &2 &3 &- &Softmax \\ \hline
CNN \cite{demir2020new}    & 6      & 3           & -           & 2           & -                                                              & Softmax                                                        \\ \hline
TFCNN \cite{mu2021environmental}   & 3      & 2           & -           & 2           & -                                                              & Softmax                                                        \\ \hline

\end{tabular}
\label{tab:blc}
\end{table*}

\begin{table*}[]
\footnotesize
\centering
\caption{Assessing the EnTL model's performance in comparison to standard baseline methods.}
\begin{tabular}{|l|c|c|c|c|c|c|c|}
\hline
\rowcolor[HTML]{C0C0C0} 
 & \textbf{Accuracy} & \textbf{MCC}  & \textbf{F1 Score} & \textbf{Precision} & \textbf{Recall} & \textbf{\begin{tabular}[c]{@{}c@{}}Kappa\\ Statistics\end{tabular}} & \textbf{\begin{tabular}[c]{@{}c@{}}ROC\\ AUC\end{tabular}} \\ \hline
\textbf{TFCNN \cite{mu2021environmental}}                                                         & 77.82\%           & 0.71          & 0.78              & 0.79               & 0.78            & 0.70                                                                & 0.85             \\ \hline
\textbf{CNN \cite{demir2020new}}                                                         & 80.45\%           & 0.76          & 0.81              & 0.88               & 0.80            & 0.74                                                                & 0.87             \\ \hline
\textbf{Dilated CNN \cite{chen2019environmental}}                                                         & 86.57\%           & 0.82          & 0.87              & 0.88               & 0.86            & 0.82                                                                & 0.91             \\ \hline
\textbf{SBCNN \cite{salamon2017deep}}                                                         & 75.23\%              & 0.68          & 0.74              & 0.79               & 0.75            & 0.67                                                                & 0.83             \\ \hline
\textbf{\begin{tabular}[c]{@{}l@{}}Our Proposed\\ Model (EnTL)\end{tabular}} & \textbf{96.72\%}  & \textbf{0.96} & \textbf{0.97}     & \textbf{0.97}      & \textbf{0.97}   & \textbf{0.96}                                                       & \textbf{0.98}    \\ \hline
\end{tabular}
\label{tab:blc1}
\end{table*}

\subsection{Discussion \& remarks} \label{sec:dr}

\begin{itemize}
    \item \textbf{Importance of our curated dataset:} Due to the fundamental nature of the traffic data, our dataset was the combination of several ambient noises, but still, we achieved a decent accuracy for honk classification. When we performed the same experiment with the baseline models, the overall accuracy decreased as all the baseline models were trained either in a controlled environment or ambient noise-free data only. Moreover, we have generated 22 hours of labeled data repository, which can further be used for developing several micro-services based on honk features.
    
    \item \textbf{Significance of our proposed data labeling model:} Due to the similar pattern of honk signature, the existing data labeling models failed to reach a good accuracy, whereas our proposed MAE model succeeded with 21\% more accuracy than existing models. Therefore, we can extrapolate that this proposed model is more suitable and will give better outcomes when the data pattern is very closely related.    
    
    \item \textbf{Role of honk classification models:} There may be circumstances where real-time detection and model execution are both required on some resource-constrained devices like IoT edge devices or smartphones. In this situation, we suggest using the MobileNet model because it is lightweight and doesn't significantly degrade performance (by about 6.53\% as seen in our tests). Furthermore, our proposed model EnTL outperforms other models and provides the best outcome.
    
    \item \textbf{Inference of context:} Proper honk classification facilitates us in recognizing the outdoor context of a location. It has had a big influence on researching passive sensing methods to develop several micro-services. In this research, we have presented a glimpse of spatio-temporal context of three different locations.
    
\end{itemize}

\section{Context identification} \label{sec:ci}
Location information can be learned by identifying the context patterns in road traffic. The location context can be detected using a variety of techniques. GPS data has been used in many studies for detecting location context, transportation mode detection, etc. Despite providing output with high precision, GPS comes with several drawbacks. These include tracking personal data without prior permission or knowledge, fast battery drains, and so on. The work presented in this paper has mainly concentrated on identifying the context of a location based on the movement of different types of vehicles in a locality. As shown in \ref{sec:phc}, the proposed model EnTL classifies different vehicular honks based on the vehicle types with higher accuracy, which can be an excellent indicator for understanding the location. Along with the classified honks, we can also use the SPL or sound pressure level of a location to detect location context. As a  case study, we have presented context identification based on spatio-temporal characteristics below. 


\subsection {Spatio-temporal context inference} \label{sec:stc}
\begin{figure*}
    \centering
    \includegraphics[width= \linewidth]{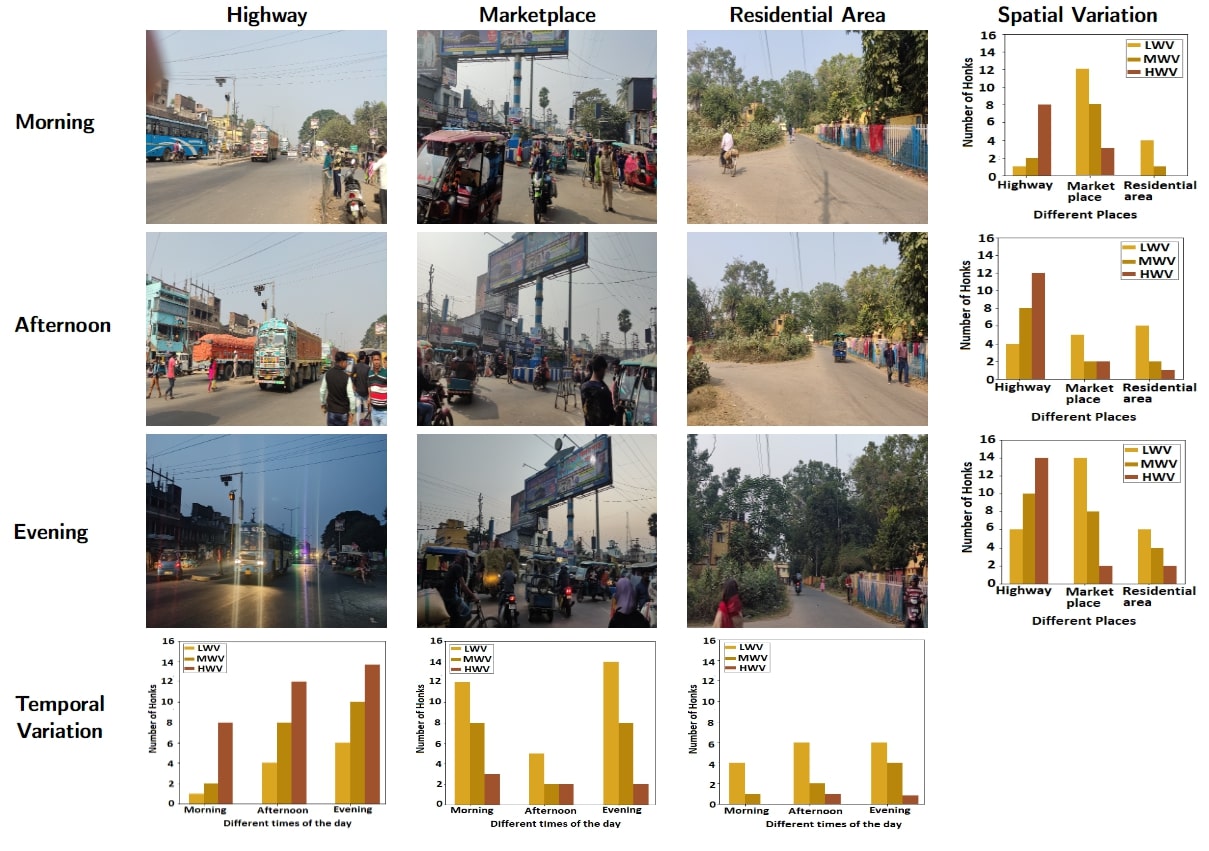}
    \caption{Variation of honking level in three different places during three different time slots}
    \label{fig:st}
\end{figure*}

To assess the context of a location, we have collected vehicular honk data from different outdoor scenes i.e. highways, market place, and residential areas for three different time slots. Continuous ten days of data have been collected for the five minutes duration of each time slot. The overall hoking signature for all the locations and time slots are depicted in Fig. \ref{fig:st}. We have also provided a snap of all the places that were visited in different time slots. From Fig. \ref{fig:st}, we have observed the following sound pattern (f\# denotes the features number for each of the locations):
\begin{itemize}
    \item Highway: [f1] Throughout the day, the highest number of honks are generated by HWV. [f2] The number of honks generated by the LWV is the lowest as compared with others. [f3] All vehicle honks increase as time proceeds. 
    \item  Market place: [f1] Honks of LWV and MWV are higher. [f3] comparatively, HWV honks are much lower. [f2] Overall honking pattern goes down in the afternoon. 
    \item Residential Area: [f1] Some LWV honks are found throughout the day [f2] HWV honks are almost negligible in this place. [f3] In comparison to other places, the number of honks is significantly less in residential areas. 
\end{itemize}
Therefore, we can say that a unique temporal pattern exists in all the locations, which distinguishes one place from another.

\subsection{Ground truth verification of context in real scenario}
\begin{figure*}
    \centering
    \includegraphics[width= 16cm, height=7cm]{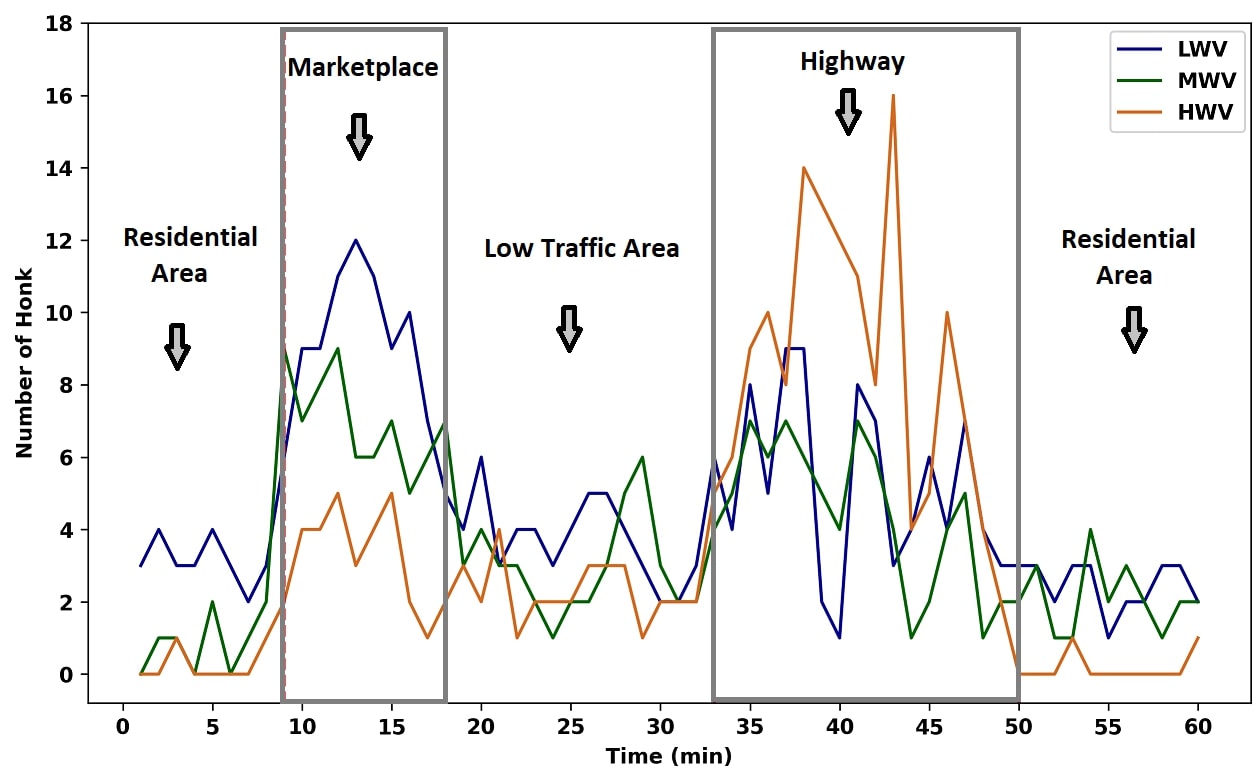}
    \caption{Ground truth verification of context considering different vehicular honking types with a physical location}
    \label{fig:gt}
\end{figure*}

\begin{figure}
    \centering
    \includegraphics[width=7cm, height= 10cm]{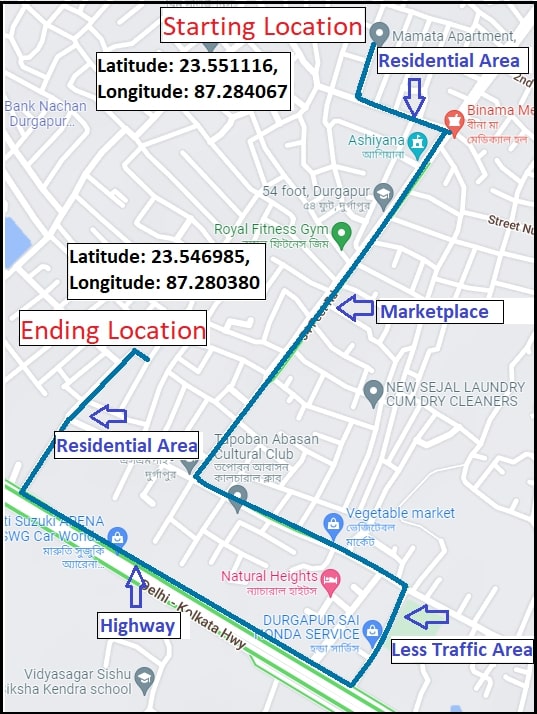}
    \caption{Street view of our physical ground truth verification route}
    \label{fig:ta}
\end{figure}

In order to verify how well our proposed system \textit{$\mathcal{A}$$\mathcal{C}$lassi$\mathcal{H}$onk} works, we moved through four different areas in a single trace and plotted the various honk data in a graph. The perceived result is presented in Fig. \ref{fig:gt} and the street view of our travel is also shown along with the latitude and longitude value in Fig. \ref{fig:ta}. We started our journey from the residential area and then passed through a marketplace, low-traffic area, high-traffic area, and again returned back to the residential area. Fig. \ref{fig:gt} shows all most similar kinds of honking patterns that we observed in Section \ref{sec:stc}. A distinguishable honking pattern among all the locations is visible, which justifies our institution to identify the context of a place.

\section{Conclusion \& future work} \label{sec:cf}
Noise pollution and vehicular honking complement each other, and their adverse impact is long-lasting for daily human life. Hence understanding the honking behavior is an important aspect based on spatio-temporal characteristics. In this paper, we have developed \textit{$\mathcal{A}$$\mathcal{C}$lassi$\mathcal{H}$onk}, a novel framework, which can classify different vehicle honks from the raw audio samples. As an application, we can detect and understand the context of a location based on the vehicle honking in road traffic. Due to the similar honking pattern of some vehicles, the existing deep learning-based data labeling models failed to obtain good accuracy in labeling the unlabeled honk samples. In this paper, we have proposed a Multi-label autoencoder (MAE), an extended AE model, that correctly labels the unlabeled honk samples for training with higher accuracy. Moreover, we have evaluated some pre-trained CNN models (ShuffleNet, MobileNet, ResNet 50, Inception V3) and proposed an ensemble model named EnTL combining four pre-trained models by using majority voting to classify vehicular honks. Experimental results illustrate that EnTL performs better than the pre-trained models and also other CNN-based baseline models with a higher percentage of accuracy. The classified honking pattern and SPL of an area are further utilized to derive the context of a location. Additionally, we have validated our model with ground truth data. 

Finally, we would like to mention a few potential future scopes and the directions of this research. In the future, we can incorporate the following features to get a more authentic context of a location: i) Our models successfully classify honks based on their vehicle types. However, the extension of honk-level features like honk duration, inter-honk gap, etc., can further provide insights into traffic conditions and the driving culture of a particular location. ii) The correlation between noise and air quality, which was studied in some existing works, can be revisited in the context of such vehicle classes. 
 iii) Our system failed to identify vehicle types when honks from multiple vehicle types overlapped in time. Suitable modification of the model is required for further improvement.

\backmatter






\noindent

\bigskip

\begin{appendices}





\end{appendices}




\bibliography{sn-bibliography}

\begin{thebibliography}{}
\providecommand{\doi}[1]{\url{https://doi.org/#1}}
\bibcommenthead

\bibitem [\protect \citeauthoryear {%
Aamir%
, Mohd~Nawi%
, Wahid%
\BCBL {}\ \BBA {} Mahdin%
}{%
Aamir%
\ \protect \BOthers {.}}{%
{\protect \APACyear {2021}}%
}]{%
aamir2021deep}
\APACinsertmetastar {%
aamir2021deep}%
\begin{APACrefauthors}%
Aamir, M.%
, Mohd~Nawi, N.%
, Wahid, F.%
\BCBL {} Mahdin, H.%
\end{APACrefauthors}%
\unskip\
\newblock
\APACrefYearMonthDay{2021}{}{}.
\newblock
{\BBOQ}\APACrefatitle {A deep contractive autoencoder for solving multiclass classification problems} {A deep contractive autoencoder for solving multiclass classification problems}.{\BBCQ}
\newblock
\APACjournalVolNumPages{Evolutionary Intelligence}{14}{}{1619--1633}.
\newblock

\newblock

\PrintBackRefs{\CurrentBib}

\bibitem [\protect \citeauthoryear {%
Abdoli%
, Cardinal%
\BCBL {}\ \BBA {} Koerich%
}{%
Abdoli%
\ \protect \BOthers {.}}{%
{\protect \APACyear {2019}}%
}]{%
abdoli2019end}
\APACinsertmetastar {%
abdoli2019end}%
\begin{APACrefauthors}%
Abdoli, S.%
, Cardinal, P.%
\BCBL {} Koerich, A.L.%
\end{APACrefauthors}%
\unskip\
\newblock
\APACrefYearMonthDay{2019}{}{}.
\newblock
{\BBOQ}\APACrefatitle {End-to-end environmental sound classification using a 1D convolutional neural network} {End-to-end environmental sound classification using a 1d convolutional neural network}.{\BBCQ}
\newblock
\APACjournalVolNumPages{Expert Systems with Applications}{136}{}{252--263}.
\newblock

\newblock

\PrintBackRefs{\CurrentBib}

\bibitem [\protect \citeauthoryear {%
Ahmed%
, Robin%
, Shafin%
\BCBL {}\ \protect \BOthers {.}}{%
Ahmed%
\ \protect \BOthers {.}}{%
{\protect \APACyear {2020}}%
}]{%
ahmed2020automatic}
\APACinsertmetastar {%
ahmed2020automatic}%
\begin{APACrefauthors}%
Ahmed, M.%
, Robin, T.I.%
, Shafin, A.A.%
\BCBL {}\ \BOthersPeriod {.}\end{APACrefauthors}%
\unskip\
\newblock
\APACrefYearMonthDay{2020}{}{}.
\newblock
{\BBOQ}\APACrefatitle {Automatic environmental sound recognition (AESR) using convolutional neural network} {Automatic environmental sound recognition (aesr) using convolutional neural network}.{\BBCQ}
\newblock
\APACjournalVolNumPages{International Journal of Modern Education \& Computer Science}{12}{5}{}.
\newblock

\newblock

\PrintBackRefs{\CurrentBib}

\bibitem [\protect \citeauthoryear {%
Allen%
\ \protect \BOthers {.}}{%
Allen%
\ \protect \BOthers {.}}{%
{\protect \APACyear {2009}}%
}]{%
allen2009spatial}
\APACinsertmetastar {%
allen2009spatial}%
\begin{APACrefauthors}%
Allen, R.W.%
, Davies, H.%
, Cohen, M.A.%
, Mallach, G.%
, Kaufman, J.D.%
\BCBL {} Adar, S.D.%
\end{APACrefauthors}%
\unskip\
\newblock
\APACrefYearMonthDay{2009}{}{}.
\newblock
{\BBOQ}\APACrefatitle {The spatial relationship between traffic-generated air pollution and noise in 2 US cities} {The spatial relationship between traffic-generated air pollution and noise in 2 us cities}.{\BBCQ}
\newblock
\APACjournalVolNumPages{Environmental research}{109}{3}{334--342}.
\newblock

\newblock

\PrintBackRefs{\CurrentBib}

\bibitem [\protect \citeauthoryear {%
Amin%
, Hassan%
\BCBL {}\ \BBA {} Jaafar%
}{%
Amin%
\ \protect \BOthers {.}}{%
{\protect \APACyear {2020}}%
}]{%
amin2020semi}
\APACinsertmetastar {%
amin2020semi}%
\begin{APACrefauthors}%
Amin, I.%
, Hassan, S.%
\BCBL {} Jaafar, J.%
\end{APACrefauthors}%
\unskip\
\newblock
\APACrefYearMonthDay{2020}{}{}.
\newblock
{\BBOQ}\APACrefatitle {Semi-Supervised Learning for Limited Medical Data Using Generative Adversarial Network and Transfer Learning} {Semi-supervised learning for limited medical data using generative adversarial network and transfer learning}.{\BBCQ}
\newblock
 \APACrefbtitle {2020 International Conference on Computational Intelligence (ICCI)} {2020 international conference on computational intelligence (icci)}\ (\BPGS\ 5--10).
\PrintBackRefs{\CurrentBib}

\bibitem [\protect \citeauthoryear {%
Andrade%
, de Lima%
, Martins%
, Zannin%
\BCBL {}\ \BBA {} da~Cunha~e Silva%
}{%
Andrade%
\ \protect \BOthers {.}}{%
{\protect \APACyear {2024}}%
}]{%
andrade2024urban}
\APACinsertmetastar {%
andrade2024urban}%
\begin{APACrefauthors}%
Andrade, E.d.L.%
, de Lima, E.A.%
, Martins, A.C.G.%
, Zannin, P.H.T.%
\BCBL {} da~Cunha~e Silva, D.C.%
\end{APACrefauthors}%
\unskip\
\newblock
\APACrefYearMonthDay{2024}{}{}.
\newblock
{\BBOQ}\APACrefatitle {Urban noise assessment in hospitals: measurements and mapping in the context of the city of Sorocaba, Brazil} {Urban noise assessment in hospitals: measurements and mapping in the context of the city of sorocaba, brazil}.{\BBCQ}
\newblock
\APACjournalVolNumPages{Environmental Monitoring and Assessment}{196}{3}{267}.
\newblock

\newblock

\PrintBackRefs{\CurrentBib}

\bibitem [\protect \citeauthoryear {%
Attallah%
}{%
Attallah%
}{%
{\protect \APACyear {2023}}%
}]{%
attallah2023cercan}
\APACinsertmetastar {%
attallah2023cercan}%
\begin{APACrefauthors}%
Attallah, O.%
\end{APACrefauthors}%
\unskip\
\newblock
\APACrefYearMonthDay{2023}{}{}.
\newblock
{\BBOQ}\APACrefatitle {CerCan{\textperiodcentered} Net: Cervical Cancer Classification Model via Multi-layer Feature Ensembles of Lightweight CNNs and Transfer Learning} {Cercan{\textperiodcentered} net: Cervical cancer classification model via multi-layer feature ensembles of lightweight cnns and transfer learning}.{\BBCQ}
\newblock
\APACjournalVolNumPages{Expert Systems with Applications}{}{}{120624}.
\newblock

\newblock

\PrintBackRefs{\CurrentBib}

\bibitem [\protect \citeauthoryear {%
Banerjee%
\ \BBA {} Sinha%
}{%
Banerjee%
\ \BBA {} Sinha%
}{%
{\protect \APACyear {2012}}%
}]{%
banerjee2012two}
\APACinsertmetastar {%
banerjee2012two}%
\begin{APACrefauthors}%
Banerjee, R.%
\BCBT {}\ \BBA {} Sinha, A.%
\end{APACrefauthors}%
\unskip\
\newblock
\APACrefYearMonthDay{2012}{}{}.
\newblock
{\BBOQ}\APACrefatitle {Two stage feature extraction using modified MFCC for honk detection} {Two stage feature extraction using modified mfcc for honk detection}.{\BBCQ}
\newblock
 \APACrefbtitle {2012 International Conference on Communications, Devices and Intelligent Systems (CODIS)} {2012 international conference on communications, devices and intelligent systems (codis)}\ (\BPGS\ 97--100).
\PrintBackRefs{\CurrentBib}

\bibitem [\protect \citeauthoryear {%
Bank%
, Koenigstein%
\BCBL {}\ \BBA {} Giryes%
}{%
Bank%
\ \protect \BOthers {.}}{%
{\protect \APACyear {2020}}%
}]{%
bank2020autoencoders}
\APACinsertmetastar {%
bank2020autoencoders}%
\begin{APACrefauthors}%
Bank, D.%
, Koenigstein, N.%
\BCBL {} Giryes, R.%
\end{APACrefauthors}%
\unskip\
\newblock
\APACrefYearMonthDay{2020}{}{}.
\newblock
{\BBOQ}\APACrefatitle {Autoencoders} {Autoencoders}.{\BBCQ}
\newblock
\APACjournalVolNumPages{arXiv preprint arXiv:2003.05991}{}{}{}.
\newblock

\newblock

\PrintBackRefs{\CurrentBib}

\bibitem [\protect \citeauthoryear {%
Bello%
\ \protect \BOthers {.}}{%
Bello%
\ \protect \BOthers {.}}{%
{\protect \APACyear {2019}}%
}]{%
bello2019sonyc}
\APACinsertmetastar {%
bello2019sonyc}%
\begin{APACrefauthors}%
Bello, J.P.%
, Silva, C.%
, Nov, O.%
, Dubois, R.L.%
, Arora, A.%
, Salamon, J.%
\BDBL {}Doraiswamy, H.%
\end{APACrefauthors}%
\unskip\
\newblock
\APACrefYearMonthDay{2019}{}{}.
\newblock
{\BBOQ}\APACrefatitle {Sonyc: A system for monitoring, analyzing, and mitigating urban noise pollution} {Sonyc: A system for monitoring, analyzing, and mitigating urban noise pollution}.{\BBCQ}
\newblock
\APACjournalVolNumPages{Communications of the ACM}{62}{2}{68--77}.
\newblock

\newblock

\PrintBackRefs{\CurrentBib}

\bibitem [\protect \citeauthoryear {%
Chen%
, Guo%
, Liang%
, Wang%
\BCBL {}\ \BBA {} Qian%
}{%
Chen%
\ \protect \BOthers {.}}{%
{\protect \APACyear {2019}}%
}]{%
chen2019environmental}
\APACinsertmetastar {%
chen2019environmental}%
\begin{APACrefauthors}%
Chen, Y.%
, Guo, Q.%
, Liang, X.%
, Wang, J.%
\BCBL {} Qian, Y.%
\end{APACrefauthors}%
\unskip\
\newblock
\APACrefYearMonthDay{2019}{}{}.
\newblock
{\BBOQ}\APACrefatitle {Environmental sound classification with dilated convolutions} {Environmental sound classification with dilated convolutions}.{\BBCQ}
\newblock
\APACjournalVolNumPages{Applied Acoustics}{148}{}{123--132}.
\newblock

\newblock

\PrintBackRefs{\CurrentBib}

\bibitem [\protect \citeauthoryear {%
Chouksey%
, Kumar%
, Parida%
, Pandey%
\BCBL {}\ \BBA {} Verma%
}{%
Chouksey%
\ \protect \BOthers {.}}{%
{\protect \APACyear {2023}}%
}]{%
chouksey2023heterogeneous}
\APACinsertmetastar {%
chouksey2023heterogeneous}%
\begin{APACrefauthors}%
Chouksey, A.K.%
, Kumar, B.%
, Parida, M.%
, Pandey, A.D.%
\BCBL {} Verma, G.%
\end{APACrefauthors}%
\unskip\
\newblock
\APACrefYearMonthDay{2023}{}{}.
\newblock
{\BBOQ}\APACrefatitle {Heterogeneous road traffic noise modeling at mid-block sections of mid-sized city in India} {Heterogeneous road traffic noise modeling at mid-block sections of mid-sized city in india}.{\BBCQ}
\newblock
\APACjournalVolNumPages{Environmental Monitoring and Assessment}{195}{11}{1349}.
\newblock

\newblock

\PrintBackRefs{\CurrentBib}

\bibitem [\protect \citeauthoryear {%
Cobos%
, Antonacci%
, Alexandridis%
, Mouchtaris%
\BCBL {}\ \BBA {} Lee%
}{%
Cobos%
\ \protect \BOthers {.}}{%
{\protect \APACyear {2017}}%
}]{%
cobos2017survey}
\APACinsertmetastar {%
cobos2017survey}%
\begin{APACrefauthors}%
Cobos, M.%
, Antonacci, F.%
, Alexandridis, A.%
, Mouchtaris, A.%
\BCBL {} Lee, B.%
\end{APACrefauthors}%
\unskip\
\newblock
\APACrefYearMonthDay{2017}{}{}.
\newblock
{\BBOQ}\APACrefatitle {A survey of sound source localization methods in wireless acoustic sensor networks} {A survey of sound source localization methods in wireless acoustic sensor networks}.{\BBCQ}
\newblock
\APACjournalVolNumPages{Wireless Communications and Mobile Computing}{2017}{}{}.
\newblock

\newblock

\PrintBackRefs{\CurrentBib}

\bibitem [\protect \citeauthoryear {%
de~la Rosa%
, G{\'o}mez-Sirvent%
, S{\'a}nchez-Reolid%
, Morales%
\BCBL {}\ \BBA {} Fern{\'a}ndez-Caballero%
}{%
de~la Rosa%
\ \protect \BOthers {.}}{%
{\protect \APACyear {2022}}%
}]{%
de2022geometric}
\APACinsertmetastar {%
de2022geometric}%
\begin{APACrefauthors}%
de~la Rosa, F.L.%
, G{\'o}mez-Sirvent, J.L.%
, S{\'a}nchez-Reolid, R.%
, Morales, R.%
\BCBL {} Fern{\'a}ndez-Caballero, A.%
\end{APACrefauthors}%
\unskip\
\newblock
\APACrefYearMonthDay{2022}{}{}.
\newblock
{\BBOQ}\APACrefatitle {Geometric transformation-based data augmentation on defect classification of segmented images of semiconductor materials using a ResNet50 convolutional neural network} {Geometric transformation-based data augmentation on defect classification of segmented images of semiconductor materials using a resnet50 convolutional neural network}.{\BBCQ}
\newblock
\APACjournalVolNumPages{Expert Systems with Applications}{206}{}{117731}.
\newblock

\newblock

\PrintBackRefs{\CurrentBib}

\bibitem [\protect \citeauthoryear {%
Demir%
, Abdullah%
\BCBL {}\ \BBA {} Sengur%
}{%
Demir%
\ \protect \BOthers {.}}{%
{\protect \APACyear {2020}}%
}]{%
demir2020new}
\APACinsertmetastar {%
demir2020new}%
\begin{APACrefauthors}%
Demir, F.%
, Abdullah, D.A.%
\BCBL {} Sengur, A.%
\end{APACrefauthors}%
\unskip\
\newblock
\APACrefYearMonthDay{2020}{}{}.
\newblock
{\BBOQ}\APACrefatitle {A new deep CNN model for environmental sound classification} {A new deep cnn model for environmental sound classification}.{\BBCQ}
\newblock
\APACjournalVolNumPages{IEEE Access}{8}{}{66529--66537}.
\newblock

\newblock

\PrintBackRefs{\CurrentBib}

\bibitem [\protect \citeauthoryear {%
Dim%
, Feitosa%
, Mota%
\BCBL {}\ \BBA {} Morais%
}{%
Dim%
\ \protect \BOthers {.}}{%
{\protect \APACyear {2020}}%
}]{%
dim2020smartphone}
\APACinsertmetastar {%
dim2020smartphone}%
\begin{APACrefauthors}%
Dim, C.A.%
, Feitosa, R.M.%
, Mota, M.P.%
\BCBL {} Morais, J.M.d.%
\end{APACrefauthors}%
\unskip\
\newblock
\APACrefYearMonthDay{2020}{}{}.
\newblock
{\BBOQ}\APACrefatitle {A Smartphone Application for Car Horn Detection to Assist Hearing-Impaired People in Driving} {A smartphone application for car horn detection to assist hearing-impaired people in driving}.{\BBCQ}
\newblock
 \APACrefbtitle {International Conference on Computational Science and Its Applications} {International conference on computational science and its applications}\ (\BPGS\ 104--116).
\PrintBackRefs{\CurrentBib}

\bibitem [\protect \citeauthoryear {%
Elhassan%
, Huang%
, Yang%
\BCBL {}\ \BBA {} Munea%
}{%
Elhassan%
\ \protect \BOthers {.}}{%
{\protect \APACyear {2021}}%
}]{%
elhassan2021dsanet}
\APACinsertmetastar {%
elhassan2021dsanet}%
\begin{APACrefauthors}%
Elhassan, M.A.%
, Huang, C.%
, Yang, C.%
\BCBL {} Munea, T.L.%
\end{APACrefauthors}%
\unskip\
\newblock
\APACrefYearMonthDay{2021}{}{}.
\newblock
{\BBOQ}\APACrefatitle {DSANet: Dilated spatial attention for real-time semantic segmentation in urban street scenes} {Dsanet: Dilated spatial attention for real-time semantic segmentation in urban street scenes}.{\BBCQ}
\newblock
\APACjournalVolNumPages{Expert Systems with Applications}{183}{}{115090}.
\newblock

\newblock

\PrintBackRefs{\CurrentBib}

\bibitem [\protect \citeauthoryear {%
Ezhilarasi%
, Sripriya%
, Suganya%
\BCBL {}\ \BBA {} Vinodhini%
}{%
Ezhilarasi%
\ \protect \BOthers {.}}{%
{\protect \APACyear {2017}}%
}]{%
ezhilarasi2017system}
\APACinsertmetastar {%
ezhilarasi2017system}%
\begin{APACrefauthors}%
Ezhilarasi, L.%
, Sripriya, K.%
, Suganya, A.%
\BCBL {} Vinodhini, K.%
\end{APACrefauthors}%
\unskip\
\newblock
\APACrefYearMonthDay{2017}{}{}.
\newblock
{\BBOQ}\APACrefatitle {A system for monitoring air and sound pollution using arduino controller with iot technology} {A system for monitoring air and sound pollution using arduino controller with iot technology}.{\BBCQ}
\newblock
\APACjournalVolNumPages{International Research Journal in Advanced Engineering and Technology (IRJAET)}{3}{2}{1781--1785}.
\newblock

\newblock

\PrintBackRefs{\CurrentBib}

\bibitem [\protect \citeauthoryear {%
Firdaus%
\ \BBA {} Ahmad%
}{%
Firdaus%
\ \BBA {} Ahmad%
}{%
{\protect \APACyear {2010}}%
}]{%
firdaus2010noise}
\APACinsertmetastar {%
firdaus2010noise}%
\begin{APACrefauthors}%
Firdaus, G.%
\BCBT {}\ \BBA {} Ahmad, A.%
\end{APACrefauthors}%
\unskip\
\newblock
\APACrefYearMonthDay{2010}{}{}.
\newblock
{\BBOQ}\APACrefatitle {Noise pollution and human health: a case study of municipal corporation of Delhi} {Noise pollution and human health: a case study of municipal corporation of delhi}.{\BBCQ}
\newblock
\APACjournalVolNumPages{Indoor and built environment}{19}{6}{648--656}.
\newblock

\newblock

\PrintBackRefs{\CurrentBib}

\bibitem [\protect \citeauthoryear {%
Garg%
, Soni%
, Saxena%
\BCBL {}\ \BBA {} Maji%
}{%
Garg%
\ \protect \BOthers {.}}{%
{\protect \APACyear {2015}}%
}]{%
garg2015applications}
\APACinsertmetastar {%
garg2015applications}%
\begin{APACrefauthors}%
Garg, N.%
, Soni, K.%
, Saxena, T.%
\BCBL {} Maji, S.%
\end{APACrefauthors}%
\unskip\
\newblock
\APACrefYearMonthDay{2015}{}{}.
\newblock
{\BBOQ}\APACrefatitle {Applications of Autoregressive integrated moving average (ARIMA) approach in time-series prediction of traffic noise pollution} {Applications of autoregressive integrated moving average (arima) approach in time-series prediction of traffic noise pollution}.{\BBCQ}
\newblock
\APACjournalVolNumPages{Noise Control Engineering Journal}{63}{2}{182--194}.
\newblock

\newblock

\PrintBackRefs{\CurrentBib}

\bibitem [\protect \citeauthoryear {%
Grondin%
\ \BBA {} Michaud%
}{%
Grondin%
\ \BBA {} Michaud%
}{%
{\protect \APACyear {2019}}%
}]{%
grondin2019lightweight}
\APACinsertmetastar {%
grondin2019lightweight}%
\begin{APACrefauthors}%
Grondin, F.%
\BCBT {}\ \BBA {} Michaud, F.%
\end{APACrefauthors}%
\unskip\
\newblock
\APACrefYearMonthDay{2019}{}{}.
\newblock
{\BBOQ}\APACrefatitle {Lightweight and optimized sound source localization and tracking methods for open and closed microphone array configurations} {Lightweight and optimized sound source localization and tracking methods for open and closed microphone array configurations}.{\BBCQ}
\newblock
\APACjournalVolNumPages{Robotics and Autonomous Systems}{113}{}{63--80}.
\newblock

\newblock

\PrintBackRefs{\CurrentBib}

\bibitem [\protect \citeauthoryear {%
Guarnaccia%
, Mastorakis%
, Quartieri%
, Tepedino%
\BCBL {}\ \BBA {} Kaminaris%
}{%
Guarnaccia%
\ \protect \BOthers {.}}{%
{\protect \APACyear {2017}}%
}]{%
guarnaccia2017development}
\APACinsertmetastar {%
guarnaccia2017development}%
\begin{APACrefauthors}%
Guarnaccia, C.%
, Mastorakis, N.E.%
, Quartieri, J.%
, Tepedino, C.%
\BCBL {} Kaminaris, S.D.%
\end{APACrefauthors}%
\unskip\
\newblock
\APACrefYearMonthDay{2017}{}{}.
\newblock
{\BBOQ}\APACrefatitle {Development of seasonal ARIMA models for traffic noise forecasting} {Development of seasonal arima models for traffic noise forecasting}.{\BBCQ}
\newblock
 \APACrefbtitle {MATEC Web of Conferences} {Matec web of conferences}\ (\BVOL~125, \BPG~05013).
\PrintBackRefs{\CurrentBib}

\bibitem [\protect \citeauthoryear {%
Gupta%
, Gupta%
, Jain%
\BCBL {}\ \BBA {} Gupta%
}{%
Gupta%
\ \protect \BOthers {.}}{%
{\protect \APACyear {2018}}%
}]{%
gupta2018noise}
\APACinsertmetastar {%
gupta2018noise}%
\begin{APACrefauthors}%
Gupta, A.%
, Gupta, A.%
, Jain, K.%
\BCBL {} Gupta, S.%
\end{APACrefauthors}%
\unskip\
\newblock
\APACrefYearMonthDay{2018}{}{}.
\newblock
{\BBOQ}\APACrefatitle {Noise pollution and impact on children health} {Noise pollution and impact on children health}.{\BBCQ}
\newblock
\APACjournalVolNumPages{The Indian Journal of Pediatrics}{85}{4}{300--306}.
\newblock

\newblock

\PrintBackRefs{\CurrentBib}

\bibitem [\protect \citeauthoryear {%
Guzhov%
, Raue%
, Hees%
\BCBL {}\ \BBA {} Dengel%
}{%
Guzhov%
\ \protect \BOthers {.}}{%
{\protect \APACyear {2021}}%
}]{%
guzhov2021esresnet}
\APACinsertmetastar {%
guzhov2021esresnet}%
\begin{APACrefauthors}%
Guzhov, A.%
, Raue, F.%
, Hees, J.%
\BCBL {} Dengel, A.%
\end{APACrefauthors}%
\unskip\
\newblock
\APACrefYearMonthDay{2021}{}{}.
\newblock
{\BBOQ}\APACrefatitle {Esresnet: Environmental sound classification based on visual domain models} {Esresnet: Environmental sound classification based on visual domain models}.{\BBCQ}
\newblock
 \APACrefbtitle {2020 25th International Conference on Pattern Recognition (ICPR)} {2020 25th international conference on pattern recognition (icpr)}\ (\BPGS\ 4933--4940).
\PrintBackRefs{\CurrentBib}

\bibitem [\protect \citeauthoryear {%
Hammer%
, Swinburn%
\BCBL {}\ \BBA {} Neitzel%
}{%
Hammer%
\ \protect \BOthers {.}}{%
{\protect \APACyear {2014}}%
}]{%
hammer2014environmental}
\APACinsertmetastar {%
hammer2014environmental}%
\begin{APACrefauthors}%
Hammer, M.S.%
, Swinburn, T.K.%
\BCBL {} Neitzel, R.L.%
\end{APACrefauthors}%
\unskip\
\newblock
\APACrefYearMonthDay{2014}{}{}.
\newblock
{\BBOQ}\APACrefatitle {Environmental noise pollution in the United States: developing an effective public health response} {Environmental noise pollution in the united states: developing an effective public health response}.{\BBCQ}
\newblock
\APACjournalVolNumPages{Environmental health perspectives}{122}{2}{115--119}.
\newblock

\newblock

\PrintBackRefs{\CurrentBib}

\bibitem [\protect \citeauthoryear {%
Hu%
, Wu%
\BCBL {}\ \BBA {} Bian%
}{%
Hu%
\ \protect \BOthers {.}}{%
{\protect \APACyear {2022}}%
}]{%
hu2022comprehensive}
\APACinsertmetastar {%
hu2022comprehensive}%
\begin{APACrefauthors}%
Hu, Q.%
, Wu, X.%
\BCBL {} Bian, L.%
\end{APACrefauthors}%
\unskip\
\newblock
\APACrefYearMonthDay{2022}{}{}.
\newblock
{\BBOQ}\APACrefatitle {Comprehensive diagnosis model of environmental impact caused by expressway vehicle emission} {Comprehensive diagnosis model of environmental impact caused by expressway vehicle emission}.{\BBCQ}
\newblock
\APACjournalVolNumPages{Environmental Monitoring and Assessment}{194}{11}{796}.
\newblock

\newblock

\PrintBackRefs{\CurrentBib}

\bibitem [\protect \citeauthoryear {%
Jariwala%
, Syed%
, Pandya%
\BCBL {}\ \BBA {} Gajera%
}{%
Jariwala%
\ \protect \BOthers {.}}{%
{\protect \APACyear {2017}}%
}]{%
jariwala2017noise}
\APACinsertmetastar {%
jariwala2017noise}%
\begin{APACrefauthors}%
Jariwala, H.J.%
, Syed, H.S.%
, Pandya, M.J.%
\BCBL {} Gajera, Y.M.%
\end{APACrefauthors}%
\unskip\
\newblock
\APACrefYearMonthDay{2017}{}{}.
\newblock
{\BBOQ}\APACrefatitle {Noise pollution \& human health: a review} {Noise pollution \& human health: a review}.{\BBCQ}
\newblock
\APACjournalVolNumPages{Indoor Built Environ}{}{}{1--4}.
\newblock

\newblock

\PrintBackRefs{\CurrentBib}

\bibitem [\protect \citeauthoryear {%
Jezdovi{\'c}%
, Popovi{\'c}%
, Radenkovi{\'c}%
, Labus%
\BCBL {}\ \BBA {} Bogdanovi{\'c}%
}{%
Jezdovi{\'c}%
\ \protect \BOthers {.}}{%
{\protect \APACyear {2021}}%
}]{%
jezdovic2021crowdsensing}
\APACinsertmetastar {%
jezdovic2021crowdsensing}%
\begin{APACrefauthors}%
Jezdovi{\'c}, I.%
, Popovi{\'c}, S.%
, Radenkovi{\'c}, M.%
, Labus, A.%
\BCBL {} Bogdanovi{\'c}, Z.%
\end{APACrefauthors}%
\unskip\
\newblock
\APACrefYearMonthDay{2021}{}{}.
\newblock
{\BBOQ}\APACrefatitle {A crowdsensing platform for real-time monitoring and analysis of noise pollution in smart cities} {A crowdsensing platform for real-time monitoring and analysis of noise pollution in smart cities}.{\BBCQ}
\newblock
\APACjournalVolNumPages{Sustainable Computing: Informatics and Systems}{31}{}{100588}.
\newblock

\newblock

\PrintBackRefs{\CurrentBib}

\bibitem [\protect \citeauthoryear {%
Kalawapudi%
, Singh%
, Dey%
, Vijay%
\BCBL {}\ \BBA {} Kumar%
}{%
Kalawapudi%
\ \protect \BOthers {.}}{%
{\protect \APACyear {2020}}%
}]{%
kalawapudi2020noise}
\APACinsertmetastar {%
kalawapudi2020noise}%
\begin{APACrefauthors}%
Kalawapudi, K.%
, Singh, T.%
, Dey, J.%
, Vijay, R.%
\BCBL {} Kumar, R.%
\end{APACrefauthors}%
\unskip\
\newblock
\APACrefYearMonthDay{2020}{}{}.
\newblock
{\BBOQ}\APACrefatitle {Noise pollution in Mumbai Metropolitan Region (MMR): An emerging environmental threat} {Noise pollution in mumbai metropolitan region (mmr): An emerging environmental threat}.{\BBCQ}
\newblock
\APACjournalVolNumPages{Environmental monitoring and assessment}{192}{}{1--20}.
\newblock

\newblock

\PrintBackRefs{\CurrentBib}

\bibitem [\protect \citeauthoryear {%
Khamparia%
\ \protect \BOthers {.}}{%
Khamparia%
\ \protect \BOthers {.}}{%
{\protect \APACyear {2019}}%
}]{%
khamparia2019sound}
\APACinsertmetastar {%
khamparia2019sound}%
\begin{APACrefauthors}%
Khamparia, A.%
, Gupta, D.%
, Nguyen, N.G.%
, Khanna, A.%
, Pandey, B.%
\BCBL {} Tiwari, P.%
\end{APACrefauthors}%
\unskip\
\newblock
\APACrefYearMonthDay{2019}{}{}.
\newblock
{\BBOQ}\APACrefatitle {Sound classification using convolutional neural network and tensor deep stacking network} {Sound classification using convolutional neural network and tensor deep stacking network}.{\BBCQ}
\newblock
\APACjournalVolNumPages{IEEE Access}{7}{}{7717--7727}.
\newblock

\newblock

\PrintBackRefs{\CurrentBib}

\bibitem [\protect \citeauthoryear {%
Khan%
}{%
Khan%
}{%
{\protect \APACyear {2022}}%
}]{%
khan2022semi}
\APACinsertmetastar {%
khan2022semi}%
\begin{APACrefauthors}%
Khan, N.%
\end{APACrefauthors}%
\unskip\
\newblock
\APACrefYearMonthDay{2022}{}{}.
\newblock
{\BBOQ}\APACrefatitle {Semi-Supervised Generative Adversarial Network for Stress Detection Using Partially Labeled Physiological Data} {Semi-supervised generative adversarial network for stress detection using partially labeled physiological data}.{\BBCQ}
\newblock
\APACjournalVolNumPages{arXiv preprint arXiv:2206.14976}{}{}{}.
\newblock

\newblock

\PrintBackRefs{\CurrentBib}

\bibitem [\protect \citeauthoryear {%
Law%
\ \BBA {} Ghosh%
}{%
Law%
\ \BBA {} Ghosh%
}{%
{\protect \APACyear {2019}}%
}]{%
law2019multi}
\APACinsertmetastar {%
law2019multi}%
\begin{APACrefauthors}%
Law, A.%
\BCBT {}\ \BBA {} Ghosh, A.%
\end{APACrefauthors}%
\unskip\
\newblock
\APACrefYearMonthDay{2019}{}{}.
\newblock
{\BBOQ}\APACrefatitle {Multi-label classification using a cascade of stacked autoencoder and extreme learning machines} {Multi-label classification using a cascade of stacked autoencoder and extreme learning machines}.{\BBCQ}
\newblock
\APACjournalVolNumPages{Neurocomputing}{358}{}{222--234}.
\newblock

\newblock

\PrintBackRefs{\CurrentBib}

\bibitem [\protect \citeauthoryear {%
Maity%
, Alim%
, Bhattacharjee%
\BCBL {}\ \BBA {} Nandi%
}{%
Maity%
, Alim%
\BCBL {}\ \protect \BOthers {.}}{%
{\protect \APACyear {2022}}%
}]{%
maity2022dehonk}
\APACinsertmetastar {%
maity2022dehonk}%
\begin{APACrefauthors}%
Maity, B.%
, Alim, A.%
, Bhattacharjee, S.%
\BCBL {} Nandi, S.%
\end{APACrefauthors}%
\unskip\
\newblock
\APACrefYearMonthDay{2022}{}{}.
\newblock
{\BBOQ}\APACrefatitle {DeHonk: A deep learning based system to characterize vehicular honks in presence of ambient noise} {Dehonk: A deep learning based system to characterize vehicular honks in presence of ambient noise}.{\BBCQ}
\newblock
\APACjournalVolNumPages{Pervasive and Mobile Computing}{88}{}{101727}.
\newblock

\newblock

\PrintBackRefs{\CurrentBib}

\bibitem [\protect \citeauthoryear {%
Maity%
, Polapragada%
, Bhattacharjee%
\BCBL {}\ \BBA {} Nandi%
}{%
Maity%
, Polapragada%
\BCBL {}\ \protect \BOthers {.}}{%
{\protect \APACyear {2022}}%
}]{%
maity2022coan}
\APACinsertmetastar {%
maity2022coan}%
\begin{APACrefauthors}%
Maity, B.%
, Polapragada, Y.%
, Bhattacharjee, S.%
\BCBL {} Nandi, S.%
\end{APACrefauthors}%
\unskip\
\newblock
\APACrefYearMonthDay{2022}{}{}.
\newblock
{\BBOQ}\APACrefatitle {CoAN: A system framework correlating the air and noise pollution sensor data} {Coan: A system framework correlating the air and noise pollution sensor data}.{\BBCQ}
\newblock
\APACjournalVolNumPages{Pervasive and Mobile Computing}{81}{}{101546}.
\newblock

\newblock

\PrintBackRefs{\CurrentBib}

\bibitem [\protect \citeauthoryear {%
Maity%
, Trinath%
, Bhattacharjee%
\BCBL {}\ \BBA {} Nandi%
}{%
Maity%
, Trinath%
\BCBL {}\ \protect \BOthers {.}}{%
{\protect \APACyear {2022}}%
}]{%
maity2022predhonk}
\APACinsertmetastar {%
maity2022predhonk}%
\begin{APACrefauthors}%
Maity, B.%
, Trinath, M.A.S.L.P.%
, Bhattacharjee, S.%
\BCBL {} Nandi, S.%
\end{APACrefauthors}%
\unskip\
\newblock
\APACrefYearMonthDay{2022}{}{}.
\newblock
{\BBOQ}\APACrefatitle {PredHonk: A Framework to Predict Vehicular Honk Count using Deep Learning Models} {Predhonk: A framework to predict vehicular honk count using deep learning models}.{\BBCQ}
\newblock
 \APACrefbtitle {TENCON 2022-2022 IEEE Region 10 Conference (TENCON)} {Tencon 2022-2022 ieee region 10 conference (tencon)}\ (\BPGS\ 1--6).
\PrintBackRefs{\CurrentBib}

\bibitem [\protect \citeauthoryear {%
Mann%
\ \BBA {} Singh%
}{%
Mann%
\ \BBA {} Singh%
}{%
{\protect \APACyear {2024}}%
}]{%
mann2024random}
\APACinsertmetastar {%
mann2024random}%
\begin{APACrefauthors}%
Mann, S.%
\BCBT {}\ \BBA {} Singh, G.%
\end{APACrefauthors}%
\unskip\
\newblock
\APACrefYearMonthDay{2024}{}{}.
\newblock
{\BBOQ}\APACrefatitle {Random effect generalized linear model-based predictive modelling of traffic noise} {Random effect generalized linear model-based predictive modelling of traffic noise}.{\BBCQ}
\newblock
\APACjournalVolNumPages{Environmental Monitoring and Assessment}{196}{2}{168}.
\newblock

\newblock

\PrintBackRefs{\CurrentBib}

\bibitem [\protect \citeauthoryear {%
Marwah%
\ \BBA {} Agrawala%
}{%
Marwah%
\ \BBA {} Agrawala%
}{%
{\protect \APACyear {2022}}%
}]{%
marwah2022covid}
\APACinsertmetastar {%
marwah2022covid}%
\begin{APACrefauthors}%
Marwah, M.%
\BCBT {}\ \BBA {} Agrawala, P.K.%
\end{APACrefauthors}%
\unskip\
\newblock
\APACrefYearMonthDay{2022}{}{}.
\newblock
{\BBOQ}\APACrefatitle {COVID-19 lockdown and environmental pollution: an Indian multi-state investigation} {Covid-19 lockdown and environmental pollution: an indian multi-state investigation}.{\BBCQ}
\newblock
\APACjournalVolNumPages{Environmental Monitoring and Assessment}{194}{2}{49}.
\newblock

\newblock

\PrintBackRefs{\CurrentBib}

\bibitem [\protect \citeauthoryear {%
Medina-Salgado%
, Sanchez-DelaCruz%
, Pozos-Parra%
\BCBL {}\ \BBA {} Sierra%
}{%
Medina-Salgado%
\ \protect \BOthers {.}}{%
{\protect \APACyear {2022}}%
}]{%
medina2022urban}
\APACinsertmetastar {%
medina2022urban}%
\begin{APACrefauthors}%
Medina-Salgado, B.%
, Sanchez-DelaCruz, E.%
, Pozos-Parra, P.%
\BCBL {} Sierra, J.E.%
\end{APACrefauthors}%
\unskip\
\newblock
\APACrefYearMonthDay{2022}{}{}.
\newblock
{\BBOQ}\APACrefatitle {Urban traffic flow prediction techniques: A review} {Urban traffic flow prediction techniques: A review}.{\BBCQ}
\newblock
\APACjournalVolNumPages{Sustainable Computing: Informatics and Systems}{35}{}{100739}.
\newblock

\newblock

\PrintBackRefs{\CurrentBib}

\bibitem [\protect \citeauthoryear {%
Mesaros%
\ \protect \BOthers {.}}{%
Mesaros%
\ \protect \BOthers {.}}{%
{\protect \APACyear {2019}}%
}]{%
mesaros2019sound}
\APACinsertmetastar {%
mesaros2019sound}%
\begin{APACrefauthors}%
Mesaros, A.%
, Diment, A.%
, Elizalde, B.%
, Heittola, T.%
, Vincent, E.%
, Raj, B.%
\BCBL {} Virtanen, T.%
\end{APACrefauthors}%
\unskip\
\newblock
\APACrefYearMonthDay{2019}{}{}.
\newblock
{\BBOQ}\APACrefatitle {Sound event detection in the DCASE 2017 challenge} {Sound event detection in the dcase 2017 challenge}.{\BBCQ}
\newblock
\APACjournalVolNumPages{IEEE/ACM Transactions on Audio, Speech, and Language Processing}{27}{6}{992--1006}.
\newblock

\newblock

\PrintBackRefs{\CurrentBib}

\bibitem [\protect \citeauthoryear {%
Michali%
, Emrouznejad%
, Dehnokhalaji%
\BCBL {}\ \BBA {} Clegg%
}{%
Michali%
\ \protect \BOthers {.}}{%
{\protect \APACyear {2021}}%
}]{%
michali2021noise}
\APACinsertmetastar {%
michali2021noise}%
\begin{APACrefauthors}%
Michali, M.%
, Emrouznejad, A.%
, Dehnokhalaji, A.%
\BCBL {} Clegg, B.%
\end{APACrefauthors}%
\unskip\
\newblock
\APACrefYearMonthDay{2021}{}{}.
\newblock
{\BBOQ}\APACrefatitle {Noise-pollution efficiency analysis of European railways: A network DEA model} {Noise-pollution efficiency analysis of european railways: A network dea model}.{\BBCQ}
\newblock
\APACjournalVolNumPages{Transportation Research Part D: Transport and Environment}{98}{}{102980}.
\newblock

\newblock

\PrintBackRefs{\CurrentBib}

\bibitem [\protect \citeauthoryear {%
Mu%
, Yin%
, Huang%
, Xu%
\BCBL {}\ \BBA {} Du%
}{%
Mu%
\ \protect \BOthers {.}}{%
{\protect \APACyear {2021}}%
}]{%
mu2021environmental}
\APACinsertmetastar {%
mu2021environmental}%
\begin{APACrefauthors}%
Mu, W.%
, Yin, B.%
, Huang, X.%
, Xu, J.%
\BCBL {} Du, Z.%
\end{APACrefauthors}%
\unskip\
\newblock
\APACrefYearMonthDay{2021}{}{}.
\newblock
{\BBOQ}\APACrefatitle {Environmental sound classification using temporal-frequency attention based convolutional neural network} {Environmental sound classification using temporal-frequency attention based convolutional neural network}.{\BBCQ}
\newblock
\APACjournalVolNumPages{Scientific Reports}{11}{1}{1--14}.
\newblock

\newblock

\PrintBackRefs{\CurrentBib}

\bibitem [\protect \citeauthoryear {%
Mushtaq%
\ \BBA {} Su%
}{%
Mushtaq%
\ \BBA {} Su%
}{%
{\protect \APACyear {2020}}%
}]{%
mushtaq2020environmental}
\APACinsertmetastar {%
mushtaq2020environmental}%
\begin{APACrefauthors}%
Mushtaq, Z.%
\BCBT {}\ \BBA {} Su, S\BHBI F.%
\end{APACrefauthors}%
\unskip\
\newblock
\APACrefYearMonthDay{2020}{}{}.
\newblock
{\BBOQ}\APACrefatitle {Environmental sound classification using a regularized deep convolutional neural network with data augmentation} {Environmental sound classification using a regularized deep convolutional neural network with data augmentation}.{\BBCQ}
\newblock
\APACjournalVolNumPages{Applied Acoustics}{167}{}{107389}.
\newblock

\newblock

\PrintBackRefs{\CurrentBib}

\bibitem [\protect \citeauthoryear {%
Navarro%
, Mart{\'\i}nez-Espa{\~n}a%
, Bueno-Crespo%
, Mart{\'\i}nez%
\BCBL {}\ \BBA {} Cecilia%
}{%
Navarro%
\ \protect \BOthers {.}}{%
{\protect \APACyear {2020}}%
}]{%
navarro2020sound}
\APACinsertmetastar {%
navarro2020sound}%
\begin{APACrefauthors}%
Navarro, J.M.%
, Mart{\'\i}nez-Espa{\~n}a, R.%
, Bueno-Crespo, A.%
, Mart{\'\i}nez, R.%
\BCBL {} Cecilia, J.M.%
\end{APACrefauthors}%
\unskip\
\newblock
\APACrefYearMonthDay{2020}{}{}.
\newblock
{\BBOQ}\APACrefatitle {Sound levels forecasting in an acoustic sensor network using a deep neural network} {Sound levels forecasting in an acoustic sensor network using a deep neural network}.{\BBCQ}
\newblock
\APACjournalVolNumPages{Sensors}{20}{3}{903}.
\newblock

\newblock

\PrintBackRefs{\CurrentBib}

\bibitem [\protect \citeauthoryear {%
Palecek%
\ \BBA {} Cerny%
}{%
Palecek%
\ \BBA {} Cerny%
}{%
{\protect \APACyear {2016}}%
}]{%
palecek2016emergency}
\APACinsertmetastar {%
palecek2016emergency}%
\begin{APACrefauthors}%
Palecek, J.%
\BCBT {}\ \BBA {} Cerny, M.%
\end{APACrefauthors}%
\unskip\
\newblock
\APACrefYearMonthDay{2016}{}{}.
\newblock
{\BBOQ}\APACrefatitle {Emergency horn detection using embedded systems} {Emergency horn detection using embedded systems}.{\BBCQ}
\newblock
 \APACrefbtitle {2016 IEEE 14th International Symposium on Applied Machine Intelligence and Informatics (SAMI)} {2016 ieee 14th international symposium on applied machine intelligence and informatics (sami)}\ (\BPGS\ 257--261).
\PrintBackRefs{\CurrentBib}

\bibitem [\protect \citeauthoryear {%
Piczak%
}{%
Piczak%
}{%
{\protect \APACyear {2015}}%
}]{%
piczak2015environmental}
\APACinsertmetastar {%
piczak2015environmental}%
\begin{APACrefauthors}%
Piczak, K.J.%
\end{APACrefauthors}%
\unskip\
\newblock
\APACrefYearMonthDay{2015}{}{}.
\newblock
{\BBOQ}\APACrefatitle {Environmental sound classification with convolutional neural networks} {Environmental sound classification with convolutional neural networks}.{\BBCQ}
\newblock
\APACjournalVolNumPages{IEEE 25th International Workshop on Machine Learning for Signal Processing (MLSP)}{}{}{1--6}.
\newblock

\newblock

\PrintBackRefs{\CurrentBib}

\bibitem [\protect \citeauthoryear {%
Saha%
\ \protect \BOthers {.}}{%
Saha%
\ \protect \BOthers {.}}{%
{\protect \APACyear {2018}}%
}]{%
saha2018raspberry}
\APACinsertmetastar {%
saha2018raspberry}%
\begin{APACrefauthors}%
Saha, A.K.%
, Sircar, S.%
, Chatterjee, P.%
, Dutta, S.%
, Mitra, A.%
, Chatterjee, A.%
\BDBL {}Saha, H.N.%
\end{APACrefauthors}%
\unskip\
\newblock
\APACrefYearMonthDay{2018}{}{}.
\newblock
{\BBOQ}\APACrefatitle {A raspberry Pi controlled cloud based air and sound pollution monitoring system with temperature and humidity sensing} {A raspberry pi controlled cloud based air and sound pollution monitoring system with temperature and humidity sensing}.{\BBCQ}
\newblock
 \APACrefbtitle {2018 IEEE 8th Annual Computing and Communication Workshop and Conference (CCWC)} {2018 ieee 8th annual computing and communication workshop and conference (ccwc)}\ (\BPGS\ 607--611).
\PrintBackRefs{\CurrentBib}

\bibitem [\protect \citeauthoryear {%
Salamon%
\ \BBA {} Bello%
}{%
Salamon%
\ \BBA {} Bello%
}{%
{\protect \APACyear {2017}}%
}]{%
salamon2017deep}
\APACinsertmetastar {%
salamon2017deep}%
\begin{APACrefauthors}%
Salamon, J.%
\BCBT {}\ \BBA {} Bello, J.P.%
\end{APACrefauthors}%
\unskip\
\newblock
\APACrefYearMonthDay{2017}{}{}.
\newblock
{\BBOQ}\APACrefatitle {Deep convolutional neural networks and data augmentation for environmental sound classification} {Deep convolutional neural networks and data augmentation for environmental sound classification}.{\BBCQ}
\newblock
\APACjournalVolNumPages{IEEE Signal Processing Letters}{24}{3}{279--283}.
\newblock

\newblock

\PrintBackRefs{\CurrentBib}

\bibitem [\protect \citeauthoryear {%
Santini%
, Ostermaier%
\BCBL {}\ \BBA {} Vitaletti%
}{%
Santini%
\ \protect \BOthers {.}}{%
{\protect \APACyear {2008}}%
}]{%
santini2008first}
\APACinsertmetastar {%
santini2008first}%
\begin{APACrefauthors}%
Santini, S.%
, Ostermaier, B.%
\BCBL {} Vitaletti, A.%
\end{APACrefauthors}%
\unskip\
\newblock
\APACrefYearMonthDay{2008}{}{}.
\newblock
{\BBOQ}\APACrefatitle {First experiences using wireless sensor networks for noise pollution monitoring} {First experiences using wireless sensor networks for noise pollution monitoring}.{\BBCQ}
\newblock
 \APACrefbtitle {Proceedings of the workshop on Real-world wireless sensor networks} {Proceedings of the workshop on real-world wireless sensor networks}\ (\BPGS\ 61--65).
\PrintBackRefs{\CurrentBib}

\bibitem [\protect \citeauthoryear {%
Segura-Garcia%
, Felici-Castell%
, Perez-Solano%
, Cobos%
\BCBL {}\ \BBA {} Navarro%
}{%
Segura-Garcia%
\ \protect \BOthers {.}}{%
{\protect \APACyear {2014}}%
}]{%
segura2014low}
\APACinsertmetastar {%
segura2014low}%
\begin{APACrefauthors}%
Segura-Garcia, J.%
, Felici-Castell, S.%
, Perez-Solano, J.J.%
, Cobos, M.%
\BCBL {} Navarro, J.M.%
\end{APACrefauthors}%
\unskip\
\newblock
\APACrefYearMonthDay{2014}{}{}.
\newblock
{\BBOQ}\APACrefatitle {Low-cost alternatives for urban noise nuisance monitoring using wireless sensor networks} {Low-cost alternatives for urban noise nuisance monitoring using wireless sensor networks}.{\BBCQ}
\newblock
\APACjournalVolNumPages{IEEE Sensors Journal}{15}{2}{836--844}.
\newblock

\newblock

\PrintBackRefs{\CurrentBib}

\bibitem [\protect \citeauthoryear {%
Sen%
, Raman%
\BCBL {}\ \BBA {} Sharma%
}{%
Sen%
\ \protect \BOthers {.}}{%
{\protect \APACyear {2010}}%
}]{%
sen2010horn}
\APACinsertmetastar {%
sen2010horn}%
\begin{APACrefauthors}%
Sen, R.%
, Raman, B.%
\BCBL {} Sharma, P.%
\end{APACrefauthors}%
\unskip\
\newblock
\APACrefYearMonthDay{2010}{}{}.
\newblock
{\BBOQ}\APACrefatitle {Horn-ok-please} {Horn-ok-please}.{\BBCQ}
\newblock
 \APACrefbtitle {Proceedings of the 8th international conference on Mobile systems, applications, and services} {Proceedings of the 8th international conference on mobile systems, applications, and services}\ (\BPGS\ 137--150).
\PrintBackRefs{\CurrentBib}

\bibitem [\protect \citeauthoryear {%
Shekhar%
, Debadarshini%
\BCBL {}\ \BBA {} Saha%
}{%
Shekhar%
\ \protect \BOthers {.}}{%
{\protect \APACyear {2022}}%
}]{%
shekhar2022liver}
\APACinsertmetastar {%
shekhar2022liver}%
\begin{APACrefauthors}%
Shekhar, C.%
, Debadarshini, J.%
\BCBL {} Saha, S.%
\end{APACrefauthors}%
\unskip\
\newblock
\APACrefYearMonthDay{2022}{}{}.
\newblock
{\BBOQ}\APACrefatitle {LiVeR: Lightweight Vehicle Detection and Classification in Real-Time} {Liver: Lightweight vehicle detection and classification in real-time}.{\BBCQ}
\newblock
\APACjournalVolNumPages{arXiv preprint arXiv:2206.06173}{}{}{}.
\newblock

\newblock

\PrintBackRefs{\CurrentBib}

\bibitem [\protect \citeauthoryear {%
Suvorov%
, Dong%
\BCBL {}\ \BBA {} Zhukov%
}{%
Suvorov%
\ \protect \BOthers {.}}{%
{\protect \APACyear {2018}}%
}]{%
suvorov2018deep}
\APACinsertmetastar {%
suvorov2018deep}%
\begin{APACrefauthors}%
Suvorov, D.%
, Dong, G.%
\BCBL {} Zhukov, R.%
\end{APACrefauthors}%
\unskip\
\newblock
\APACrefYearMonthDay{2018}{}{}.
\newblock
{\BBOQ}\APACrefatitle {Deep residual network for sound source localization in the time domain} {Deep residual network for sound source localization in the time domain}.{\BBCQ}
\newblock
\APACjournalVolNumPages{arXiv preprint arXiv:1808.06429}{}{}{}.
\newblock

\newblock

\PrintBackRefs{\CurrentBib}

\bibitem [\protect \citeauthoryear {%
Takeuchi%
, Matsumoto%
, Takeuchi%
, Kudo%
\BCBL {}\ \BBA {} Ohnishi%
}{%
Takeuchi%
\ \protect \BOthers {.}}{%
{\protect \APACyear {2014}}%
}]{%
takeuchi2014smart}
\APACinsertmetastar {%
takeuchi2014smart}%
\begin{APACrefauthors}%
Takeuchi, K.%
, Matsumoto, T.%
, Takeuchi, Y.%
, Kudo, H.%
\BCBL {} Ohnishi, N.%
\end{APACrefauthors}%
\unskip\
\newblock
\APACrefYearMonthDay{2014}{}{}.
\newblock
{\BBOQ}\APACrefatitle {A smart-phone based system to detect warning sound for hearing impaired people} {A smart-phone based system to detect warning sound for hearing impaired people}.{\BBCQ}
\newblock
 \APACrefbtitle {International Conference on Computers for Handicapped Persons} {International conference on computers for handicapped persons}\ (\BPGS\ 506--511).
\PrintBackRefs{\CurrentBib}

\bibitem [\protect \citeauthoryear {%
Vera-Diaz%
, Pizarro%
\BCBL {}\ \BBA {} Macias-Guarasa%
}{%
Vera-Diaz%
\ \protect \BOthers {.}}{%
{\protect \APACyear {2018}}%
}]{%
vera2018towards}
\APACinsertmetastar {%
vera2018towards}%
\begin{APACrefauthors}%
Vera-Diaz, J.M.%
, Pizarro, D.%
\BCBL {} Macias-Guarasa, J.%
\end{APACrefauthors}%
\unskip\
\newblock
\APACrefYearMonthDay{2018}{}{}.
\newblock
{\BBOQ}\APACrefatitle {Towards end-to-end acoustic localization using deep learning: From audio signals to source position coordinates} {Towards end-to-end acoustic localization using deep learning: From audio signals to source position coordinates}.{\BBCQ}
\newblock
\APACjournalVolNumPages{Sensors}{18}{10}{3418}.
\newblock

\newblock

\PrintBackRefs{\CurrentBib}

\bibitem [\protect \citeauthoryear {%
Wang%
\ \protect \BOthers {.}}{%
Wang%
\ \protect \BOthers {.}}{%
{\protect \APACyear {2019}}%
}]{%
wang2019pulmonary}
\APACinsertmetastar {%
wang2019pulmonary}%
\begin{APACrefauthors}%
Wang, C.%
, Chen, D.%
, Hao, L.%
, Liu, X.%
, Zeng, Y.%
, Chen, J.%
\BCBL {} Zhang, G.%
\end{APACrefauthors}%
\unskip\
\newblock
\APACrefYearMonthDay{2019}{}{}.
\newblock
{\BBOQ}\APACrefatitle {Pulmonary image classification based on inception-v3 transfer learning model} {Pulmonary image classification based on inception-v3 transfer learning model}.{\BBCQ}
\newblock
\APACjournalVolNumPages{IEEE Access}{7}{}{146533--146541}.
\newblock

\newblock

\PrintBackRefs{\CurrentBib}

\bibitem [\protect \citeauthoryear {%
Wen%
, Guo%
\BCBL {}\ \BBA {} Li%
}{%
Wen%
\ \protect \BOthers {.}}{%
{\protect \APACyear {2023}}%
}]{%
wen2023novel}
\APACinsertmetastar {%
wen2023novel}%
\begin{APACrefauthors}%
Wen, H.%
, Guo, W.%
\BCBL {} Li, X.%
\end{APACrefauthors}%
\unskip\
\newblock
\APACrefYearMonthDay{2023}{}{}.
\newblock
{\BBOQ}\APACrefatitle {A novel deep clustering network using multi-representation autoencoder and adversarial learning for large cross-domain fault diagnosis of rolling bearings} {A novel deep clustering network using multi-representation autoencoder and adversarial learning for large cross-domain fault diagnosis of rolling bearings}.{\BBCQ}
\newblock
\APACjournalVolNumPages{Expert Systems with Applications}{225}{}{120066}.
\newblock

\newblock

\PrintBackRefs{\CurrentBib}

\bibitem [\protect \citeauthoryear {%
Wicker%
, Tyukin%
\BCBL {}\ \BBA {} Kramer%
}{%
Wicker%
\ \protect \BOthers {.}}{%
{\protect \APACyear {2016}}%
}]{%
wicker2016nonlinear}
\APACinsertmetastar {%
wicker2016nonlinear}%
\begin{APACrefauthors}%
Wicker, J.%
, Tyukin, A.%
\BCBL {} Kramer, S.%
\end{APACrefauthors}%
\unskip\
\newblock
\APACrefYearMonthDay{2016}{}{}.
\newblock
{\BBOQ}\APACrefatitle {A nonlinear label compression and transformation method for multi-label classification using autoencoders} {A nonlinear label compression and transformation method for multi-label classification using autoencoders}.{\BBCQ}
\newblock
 \APACrefbtitle {Pacific-Asia Conference on Knowledge Discovery and Data Mining} {Pacific-asia conference on knowledge discovery and data mining}\ (\BPGS\ 328--340).
\PrintBackRefs{\CurrentBib}

\bibitem [\protect \citeauthoryear {%
Zamora%
, Calafate%
, Cano%
\BCBL {}\ \BBA {} Manzoni%
}{%
Zamora%
\ \protect \BOthers {.}}{%
{\protect \APACyear {2017}}%
}]{%
zamora2017accurate}
\APACinsertmetastar {%
zamora2017accurate}%
\begin{APACrefauthors}%
Zamora, W.%
, Calafate, C.T.%
, Cano, J\BHBI C.%
\BCBL {} Manzoni, P.%
\end{APACrefauthors}%
\unskip\
\newblock
\APACrefYearMonthDay{2017}{}{}.
\newblock
{\BBOQ}\APACrefatitle {Accurate ambient noise assessment using smartphones} {Accurate ambient noise assessment using smartphones}.{\BBCQ}
\newblock
\APACjournalVolNumPages{Sensors}{17}{4}{917}.
\newblock

\newblock

\PrintBackRefs{\CurrentBib}

\bibitem [\protect \citeauthoryear {%
Zhou%
, Song%
\BCBL {}\ \BBA {} Shu%
}{%
Zhou%
\ \protect \BOthers {.}}{%
{\protect \APACyear {2017}}%
}]{%
zhou2017using}
\APACinsertmetastar {%
zhou2017using}%
\begin{APACrefauthors}%
Zhou, H.%
, Song, Y.%
\BCBL {} Shu, H.%
\end{APACrefauthors}%
\unskip\
\newblock
\APACrefYearMonthDay{2017}{}{}.
\newblock
{\BBOQ}\APACrefatitle {Using deep convolutional neural network to classify urban sounds} {Using deep convolutional neural network to classify urban sounds}.{\BBCQ}
\newblock
\APACjournalVolNumPages{IEEE Region 10 Conference(TENCON)}{}{}{3089--3092}.
\newblock

\newblock

\PrintBackRefs{\CurrentBib}

\bibitem [\protect \citeauthoryear {%
Zipf%
, Primack%
\BCBL {}\ \BBA {} Rothendler%
}{%
Zipf%
\ \protect \BOthers {.}}{%
{\protect \APACyear {2020}}%
}]{%
zipf2020citizen}
\APACinsertmetastar {%
zipf2020citizen}%
\begin{APACrefauthors}%
Zipf, L.%
, Primack, R.B.%
\BCBL {} Rothendler, M.%
\end{APACrefauthors}%
\unskip\
\newblock
\APACrefYearMonthDay{2020}{}{}.
\newblock
{\BBOQ}\APACrefatitle {Citizen scientists and university students monitor noise pollution in cities and protected areas with smartphones} {Citizen scientists and university students monitor noise pollution in cities and protected areas with smartphones}.{\BBCQ}
\newblock
\APACjournalVolNumPages{PloS one}{15}{9}{e0236785}.
\newblock

\newblock

\PrintBackRefs{\CurrentBib}

\end{thebibliography}


\end{document}